\documentclass[prb,amsmath,amssymb,twocolumn,showpacs]{revtex4-1}
\usepackage[normalem]{ulem}
\usepackage{amsmath,amssymb}
\usepackage[bookmarks=true,colorlinks,linkcolor=blue,urlcolor=blue,citecolor=blue]{hyperref}
\usepackage{graphicx,amsmath,amsfonts,dsfont}
\usepackage[usenames]{color}
\usepackage{epstopdf}
\usepackage{verbatim}
\usepackage{amsthm}
\usepackage{enumitem}
\usepackage{subfigure} %for figures in the same box
\hypersetup{colorlinks,linkcolor=blue,filecolor=green,urlcolor=blue,citecolor=blue}

\newcommand{\be}{\begin{equation}}
\newcommand{\ee}{\end{equation}}

\newcommand{\ket}[1]{\left\vert#1\right\rangle}
\newcommand{\bra}[1]{\left\langle#1\right\vert}

\newcommand{\syse}{\end{array}\right.}

\newcommand{\ag}[1]{}

\begin{document}

\title{Entanglement entropy in a periodically driven quantum Ising ring}

\author{Tony J. G. Apollaro$^{1,3}$}
\author{G. Massimo Palma$^{1}$}
\author{Jamir Marino$^{2}$}

\affiliation{$^1$NEST, Istituto Nanoscienze-CNR and Dipartimento di Fisica e Chimica, Universit$\grave{a}$  degli Studi di Palermo, via Archirafi 36, I-90123 Palermo, Italy}

 \affiliation{$^2$Institute of Theoretical Physics, University of Cologne, D-50937 Cologne, Germany}
 
 \affiliation{$^3$Centre for Theoretical Atomic, Molecular, and Optical Physics, School of Mathematics and Physics, Queen's University Belfast, BT7,1NN, United Kingdom}

\begin{abstract}
We  numerically study  the dynamics of entanglement entropy, induced by an oscillating time periodic driving of the transverse field, $h(t)$, of a  one-dimensional quantum Ising ring. We consider several realizations of $h(t)$, and we find a number of results in  analogy with entanglement entropy dynamics induced by a sudden quantum quench. After a short-time relaxation, the dynamics of entanglement entropy synchronises with $h(t)$, displaying an oscillatory behaviour at the frequency of the driving. Synchronisation in the dynamics of entanglement entropy is  spoiled  by the appearance  of quasi-revivals  which fade out in the thermodynamic limit, and which we interpret using a quasi-particle picture adapted to periodic drivings. 
We show that the  time-averaged entanglement entropy in the synchronised regime obeys a volume law scaling with the subsystem's size. Such result is reminiscent of a thermal state or  a Generalised Gibbs ensemble, although the system does not heat up towards infinite temperature as a consequence of the integrability of the model.
\end{abstract}

\pacs{75.10.Jm, 03.67.Bg}

%\date{\today}

\maketitle

The past decade has witnessed a remarkable revival of  interest for the non-equilibrium dynamics of quantum many body systems, mostly triggered by the  advances in the experimental control of the tunable coupling, as well as of the environment isolation, of ultracold atomic lattices \cite{Zwerger}.  The possibility to observe the coherent time-resolved  quantum dynamics of many body systems has stimulated considerable theoretical interest for the different dynamical regimes of many-particle quantum systems,
%\cite{Eisert,Diehl}
pioneered by studies on quantum quenches \cite{Silva,*Eisert,*Diehl}. Albeit thermalization is eventually expected in ergodic systems \cite{Deutsch, *Srednicki, *Rigol}, intermediate time dynamics can display novel features determined by the nature of the driven out of equilibrium dynamics. A paradigmatic example of this scenario is given by periodically driven many body systems \cite{Grifoni}, which can be traditionally employed  to engineer hopping in optical lattices \cite{Weiss05, *Sias07, *Zenesini09,*Struck11, *Burton13}, design artificial gauge fields in cold atoms \cite{Jacksh03, *Struck13, *Atala13, *Goldman14}, or more recently to stabilize novel topological states of matter \cite{Oka09, *Kitagawa11, *Jotzu14, *Bloch15, *Refael11, *Levin13, *Wang13, *Iadecola15, *RigolChern}. Such plethora of promising applications has stimulated a substantial theoretical debate on the properties of many-body dynamics under periodic driving \cite{Dal3, *Rigol14}. In the absence of a cooling mechanism, 
%\cite{Refael}
ergodic non-integrable systems are expected to heat up towards an infinite temperature state \cite{Refael, *Lazarides14PRE, *Ponte15, *Genske}, as a consequence of resonant energy absorption from the driving. However, even for a isolated periodically driven systems, heating can occur on an extremely long time scale, 
%\cite{DeR15}
and a richer scenario is expected at the intermediate stages of the dynamics, like, for instance, the emergence of novel pre-thermalized Floquet metastable states \cite{DeR15, *Demler15, *Abanin15}. 

At the same time, while entanglement entropy \cite{Saro,*Doyon, *0305-4470-38-13-011}  is of key importance in condensed matter to characterise topological phases %\cite{Kitaev,Wen,Jiang} 
or to probe non-equilibrium dynamics \cite{Kitaev,*Wen,*Jiang,*Daley, *PhysRevA.84.063615,*PhysRevB.91.245138}, few works have  addressed so far the evolution and the scaling of entanglement entropy in periodically driven quantum many body systems \cite{Sen, *CC, *ANDP:ANDP200810299, *1751-8121-42-50-504003, *PhysRevA.91.022318, *PhysRevA.78.012330}. In contrast, for a quantum quench protocol, the linear growth in time of entanglement entropy and its linear scaling with the system size in the asymptotic steady state, are  established results \cite{Fagotti08,  *Kollath, *DeChiara, *Igloi, *Collura,  *2016arXiv160302669P, *PhysRevB.89.104303}. 

In this work we investigate an archetypical one dimensional soluble model, the quantum Ising ring, where it has been shown that the dynamics of local observables synchronises with the driving frequency, reaching a non-trivial stationary state and where heating towards infinite temperature is prevented by the underlying integrability of the model \cite{Angelo, *Angelo2}, a circumstance shared by a larger class of exactly soluble models
 %\cite{Laz} 
 and by periodically driven interacting systems in the many body localized phase \cite{Laz, *Roy, *LazMBL}.  In the following we show that the dynamics and the scaling of entanglement entropy induced by a periodic drive and by a sudden quench share several properties -- as already anticipated in other instances of periodically driven systems \cite{Laz}. More specifically, we focus on the one dimensional quantum Ising ring (QIC) Hamiltonian, with a sinusoidal time dependent driving of the transverse magnetic field,  $h(t)=h_0+\Delta h \sin \omega t$
\begin{equation}\label{Ising}
\hat{H}(t)=-\frac{J}{2}\sum_{j=1}^N\left[\hat{\sigma}_j^x\hat{\sigma}_{j+1}^x+h(t)\hat{\sigma}_j^z\right],
\end{equation}
where $\hat{\sigma}^{\alpha}$ ($\alpha=x,y,z$) are the Pauli matrices and where periodic boundary conditions have been assumed ($\hat{\vec{\sigma}}_{N+1}=\hat{\vec{\sigma}}_1$). For $J>0$  the model is  ferromagnetic, and it exhibits, in the thermodynamical limit $N\rightarrow \infty$, an Ising-type second order quantum phase transition at $h_0=1$~\cite{Sachdev, *PhysRevA.80.032102}, which separates a ferromagnetic (FM) $h_0<1$ from a paramagnetic phase (PM) $h_0>1$. 
In the following we set $J=1$, which we adopt as our time and energy unit and,
without loss of generality, we assume $h_0\geq 0$.

The Hamiltonian given in Eq.~\ref{Ising} can be recast in a quadratic form via a Jordan-Wigner transformation 
%\cite{Sachdev} 
introducing spinless fermionic operators
$\hat{c_j}=e^{-i \pi \sum_{i=1}^{j-1}\hat{\sigma}^+_i\hat{\sigma}^-_i} \hat{\sigma}^-_j,$ where $\hat{\sigma}^{\pm}_j=\frac{\hat{\sigma}^x_i\pm i \hat{\sigma}^y_i}{2}$. After textbook algebra and Fourier transforming the spin-less fermions $c_i$, one recovers the following expression for the hamiltonian $\hat{H}(t)=\sum_{k>0}\hat{h}_k(t)$, where $\hat{h}_k(t)=\left(h(t)-\cos k\right)\left( \hat{c}_k^{\dagger}\hat{c}_k- \hat{c}_{-k}\hat{c}_{-k}^{\dagger}\right)+\sin k \left(\hat{c}_k^{\dagger}\hat{c}_{-k}^{\dagger}-\hat{c}_k\hat{c}_{-k}\right)$. Each $\hat{h}_k(t)$ acts on a two-dimensional subspace spanned by the basis $\{ \hat{c}^{\dagger}_k\hat{c}^{\dagger}_{-k}\ket{0},\ket{0} \}$, where $\ket{0}$ is the vacuum of the Jordan-Wigner fermions $\hat{c}_k$, and therefore can be written in this subspace in terms of Pauli matrices as
$\hat{h}_k(t)=\left(h(t)-\cos k\right)\hat{\sigma}^z+(\sin k)\hat{\sigma}^x$.
We consider the system initialised in the ground state of $\hat{H}(t=0)$, which assume a BCS-like form \begin{equation}\label{E.tstate}
\ket{\Psi(0)}=\prod_{k>0}\left(v_k(0)+u_k(0)\hat{c}_k^{\dagger}\hat{c}_{-k}^{\dagger}\right)\ket{0}
\end{equation}with
$\left(u_k(0),v_k(0)\right)=\left(-\sin\left(\frac{\theta_k}{2}\right),\cos\left(\frac{\theta_k}{2}\right)\right)$, $\theta_k=\tan^{-1}\frac{ \sin k}{h(0)-\cos k}$, and $k=\frac{\pi}{N},\frac{3\pi}{N},\dots,\frac{(N-1)\pi}{N}$, given by the positive parity sector subspace.
%~\cite{PhysRevA.80.032102}. 

Notice that the Hamiltonian in Eq.~\ref{Ising} can be decomposed in the direct sum of positive, $H^+$, and negative, $H^-$, parity-conserving Hamiltonians $\hat{H}=\hat{H}^++\hat{H}^-$. For every finite even $N$, the ground state belongs to the positive parity subspace (and assumes the BCS-like form)~\cite{1751-8121-47-2-025303}; in our analysis, we assumed the system prepared in this state, before applying the periodic drive.
As the dynamics induced by $\hat{H}(t)$  mixes neither subspaces with opposite parity nor with different momenta, we can restrict to the positive parity sector where the time evolved state can be written as $\ket{\Psi(t)}=\prod_{k>0}\ket{\psi_k(t)}=\prod_{k>0}\left(v_k(t)+u_k(t)\hat{c}_k^{\dagger}\hat{c}_{-k}^{\dagger}\right)\ket{0}$. Finally,  the state coefficients $\{v_k(t),u_k(t)\}$
are given by the solution of the Bogoliubov-de Gennes (BdG) equations \cite{Angelo},
\begin{equation}\label{E.BdG}
\begin{pmatrix}
\dot{u}_k(t)\\
\dot{v}_k(t)
\end{pmatrix}= -i
\begin{pmatrix}
(h(t)-\cos k) & \sin k \\
\sin k & -(h(t)-\cos k) 
\end{pmatrix}
\begin{pmatrix}
u_k(t)\\
v_k(t)
\end{pmatrix}~,
\end{equation}
for each Fourier mode $k$ with the initial condition $\{v_k(0),u_k(0)\}=\{0,1\}$.
%With the initial condition given by the ground state at $t=0$, 
%\begin{equation}\label{E.tstate}
%\ket{\Psi(0)}=\prod_{k>0}\left(v_k(0)+u_k(0)\hat{c}_k^{\dagger}\hat{c}_{-k}^{\dagger}\right)\ket{0}
%\end{equation}with
%$\left(u_k(0),v_k(0)\right)=\left(-\sin\left(\frac{\theta_k}{2}\right),\cos\left(\frac{\theta_k}{2}\right)\right)$, and $\theta_k=\tan^{-1}\frac{ \sin k}{h(0)-\cos k}$, 

Eqs.~\ref{Ising}-\ref{E.BdG} are valid for arbitrary $h(t)$, provided parity and translational invariance are conserved. If $h(t)$ is a periodic function
we can make use of the Floquet formalism~\cite{Grifoni} to determine the evolved state $\ket{\Psi(t)}$.   Throughout the paper we  assume a sinusoidal driving of the magnetic field in Eq.~\ref{Ising}, $h(t)=h+\Delta h \sin \omega t$.
 In particular, setting $t=nT+\delta t$, we have 
\begin{eqnarray}\label{E.Floquet}
\ket{\Psi(t)}&=&\hat{U}\left(t,0\right)\ket{\Psi(0)}=\prod_{k>0}\hat{U}_k(t,0)) \ket{\psi_k(0)}\\
&=&\prod_{k>0}\hat{U}_k\left(\delta t,0\right)\hat{U}_k^n\left(T,0\right)\ket{\psi_k(0)},
\end{eqnarray}
where $T=\frac{2 \pi}{\omega}$ is the period of the driving and $0<\delta t <T$.
Floquet theory allows to write
the time propagator $\hat{U}\left(t,0\right)$, over one period, through the following spectral representation in the Floquet basis
\begin{equation}\label{E.UFloquet}
\hat{U}_k(T,0)=\sum_{i=1,2} e^{-i \mu_{k_i} t}\ket{\phi_{k_i}(0)}\!\!\bra{\phi_{k_i}(0)},
\end{equation}
where $\{\mu_{k_i},\ket{\phi_{k_i}(0)}\}$ are the (two) eigenvalues and eigenvectors of $\hat{U}_k(T,0)=\mathcal{T}\left[\exp\left( -i \int_0^T\!dt\, \hat{h}_k(t)\right) \right]$ ($\mathcal{T}[...]$ is the time-ordered product). The Floquet formalism is therefore particularly useful to determine the stroboscopic dynamics at times $nT$. In particular,  from Eq.~\ref{E.Floquet}, we find
\begin{equation}\label{E.Floquetstrobo}
\ket{\Psi(nT)}=\prod_{k>0}\sum_{i=1,2} e^{-i \mu_{k_i} n T}\ket{\phi_{k_i}(0)}\!\!\bra{\phi_{k_i}(0)}\psi_k(0)\rangle.
\end{equation}
Applying such a procedure to the initial state, Eq.~(\ref{E.tstate}), we obtain the following expression for the state  at times $nT$ (see also Refs. \cite{Angelo}), 
\begin{equation}\label{E.tstrobostate}
\ket{\Psi(nT)}=\prod_{k>0}\left(v_k(nT)+u_k(nT)\hat{c}_k^{\dagger}\hat{c}_{-k}^{\dagger}\right)\ket{0}~,
\end{equation}
where the invariance of the BCS-like form of the evolved state follows from the fact that the Floquet propagator $\hat{U}\left(T,0\right)$ mixes only states with the same parity.

The Floquet formalism has been employed to show \cite{Angelo} that in the quantum Ising ring (and in all the integrable models equivalent to quadratic fermionic/bosonic Hamiltonians \cite{Laz}), observables synchronize with the driving and do not heat up towards a featureless infinite temperature state, being absorption of energy constrained by the  integrability of the model. This aspect is manifest in the long-time limit of observables, captured by a periodic Generalized Gibbs Ensemble.
% -- after an initial fade out controlled by destructive quantum interference, in full analogy with inhomogeneous dephasing occurring after a global sudden quantum quench. 
Such ensemble follows from a generalisation to periodically driven systems of the Generalized Gibbs Ensemble, which is believed to describe the asymptotic steady states of quantum quenched integrable models (for a report on the state of the art see Refs. \onlinecite{Caux, Essler}). 
Indeed, in quadratic integrable periodically driven systems  \cite{Laz}, an extensive number of time-dependent (periodic) conserved quantities $I_\alpha (t)$ (with $\alpha=1,...,N$)  can be constructed and, according to maximisation of entropy, the statistical ensemble $\rho_{PGGE}\sim \exp\left(-\sum_\alpha\lambda_\alpha I_\alpha(t)\right)$ properly describes the long time behaviour of observables, provided the Langrange multipliers  $\lambda_\alpha$ are fixed by the expectation values of the conserved charges on the initial state, $\langle \psi(0)|I_\alpha(0)|\psi(0)\rangle=\operatorname{Tr}[\rho_{PGGE}I_\alpha(0)]$. 

The main goal of our work is to study the interplay between entanglement entropy and the long-time synchronised dynamics of a periodically driven quantum Ising ring.
%We recall that the entanglement entropy, $S_l\equiv \operatorname{Tr}[\rho_l\log_2 \rho_l]$, of a subsystem of size $l$ 
%characterises the reduced density matrix $\rho_l$, 

We recall that the entanglement entropy, $S_l\equiv \operatorname{Tr}[\rho_l\log_2 \rho_l]$, of a subsystem of size $l$, quantifies the entanglement between the reduced density matrix $\rho_l$ and its complementary part $\rho_{N-l}$ for pure states of the total system. Entanglement entropy has been extensively investigated both at equilibrium and in out-of-equilibrium settings~\cite{1742-5468-2004-12-P12005, *Doyon, *FranchiniJPA08, *FranchiniJPA07,*igloi2}. It satisfies the area law when the quantum Ising ring is in the non-critical ground state \footnote{With the notable exception of the critical point where it scales logarithmically $S_l\sim \log l$, see for instance \cite{Doyon, Saro}.}, while at finite temperature or after a quantum quench it scales with the volume of the subsystem, $S_l\sim l$ (see for instance~\onlinecite{Fagotti08, Igloi}). 

\begin{figure}[!t]
{%
       \includegraphics[height=0.21\textheight,width=0.43\textwidth]{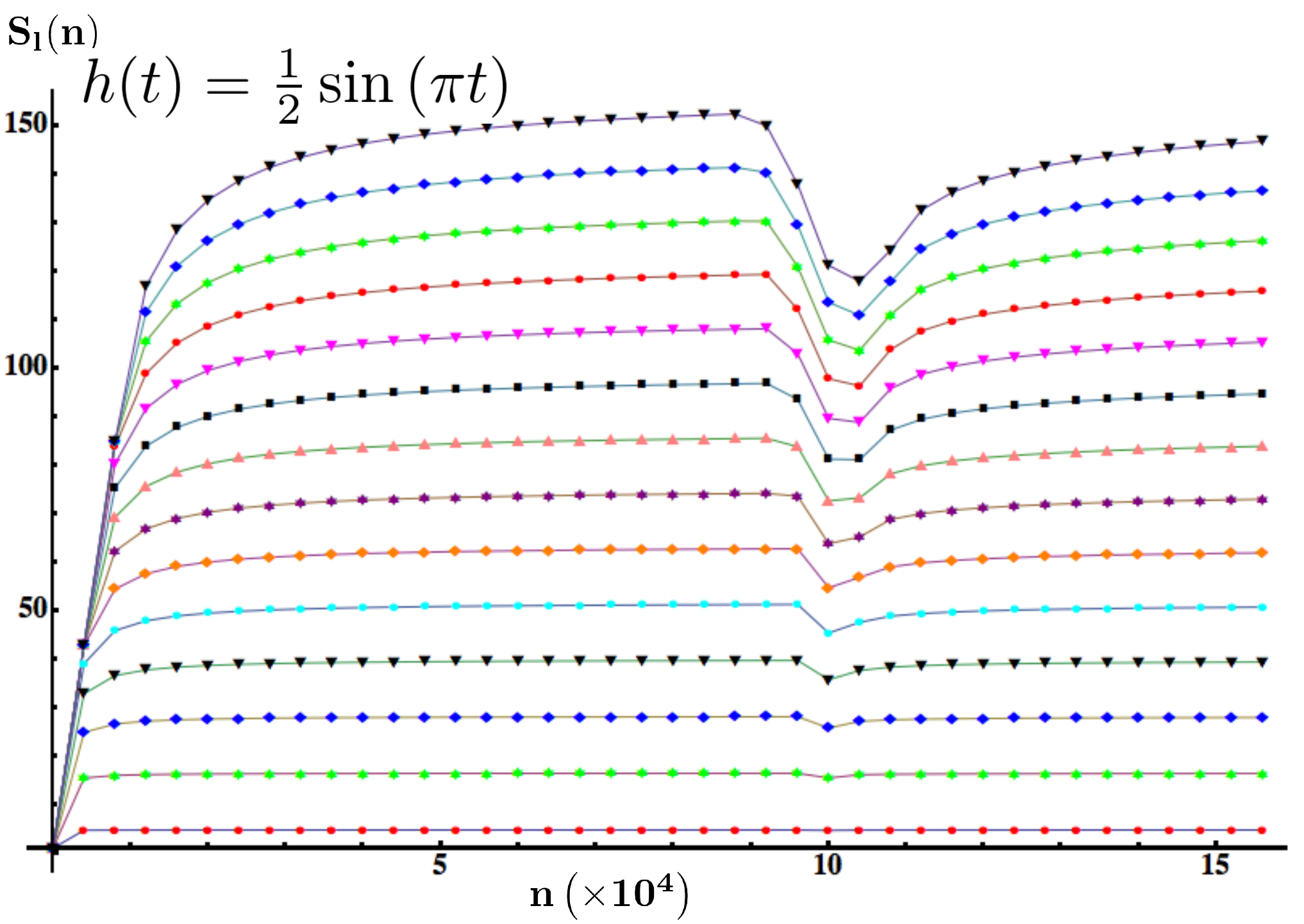}  
    }
    \hfill
{%
       \includegraphics[height=0.21\textheight,width=0.43\textwidth]{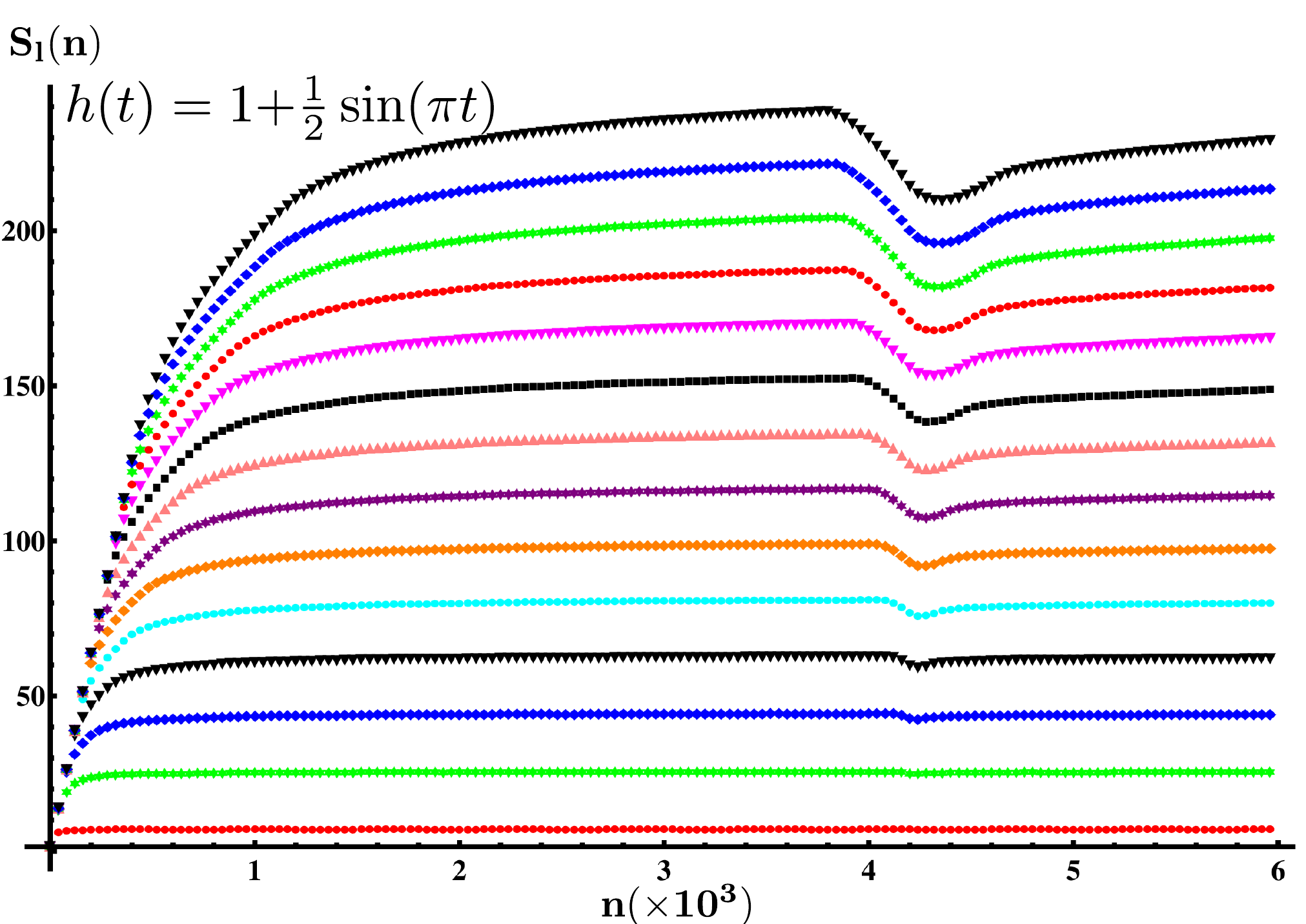}
    }
    \caption{(color online) 
%\jm{notazione scientifica lungo l'asse x; colori piu' vivi, sento ancora una differenza rispetto alle altre :)}
(upper panel) Time evolution (in units of numbers of cycles) of the entanglement entropy for driving inside the FM phase $h(t)=\frac{1}{2}\sin(\pi t)$, for a quantum Ising ring of length $N=8192$ spins. The curves correspond to subsystem's size $l=20, 80, 140,\dots,800$ (bottom-up) and the markers correspond to the stroboscopic dynamics of the block entanglement entropy $S_l(n)$.(lower panel) The same as in the upper panel for a driving around the critical point $h(t)=1+\frac{1}{2}\sin(\pi t)$. Notice the different time scales at which the common dynamical features (linear increase, quasi-revivals) occur. These differences in time scales are related to the different maximum speed of the spreading of Floquet-quasi particles. 
}
    \label{FigFM}
  \end{figure}

The non-trivial nature of the synchronised dynamics, lead us to wonder whether - after a large number $n$ of driving cycles, the scaling of $S_l$  obeys a volume law (as the analogy with quantum quenches would suggest), or some novel scaling of entanglement entropy, $S_l\sim l^s$, (with $s\neq 1$) can be observed. 

In order to compute the entanglement entropy in the periodically driven regime, we consider a bipartition made of two blocks of contiguous spins, respectively of length $\{l,N-l\}$, and we extract the correlation matrix $
\Pi=
\begin{pmatrix}
\alpha & \beta^{\dagger}\\
\beta & \mathbf{1}-\alpha
\end{pmatrix} $  where the matrix elements $\alpha_{mn}(t)=\bra{\Psi(t)}\hat{c}_m\hat{c}_n^{\dagger}\ket{\Psi(t)}$ and $\beta_{mn}(t)=\bra{\Psi(t)}\hat{c}_m\hat{c}_n\ket{\Psi(t)}$ are the $l$-dimensional correlation matrices of Jordan-Wigner fermions, $m,n=1,...,l$, and $\mathbf{1}$ is the unit matrix in the $l$ dimensional matrix space (for technical details, see \cite{Vidal2003, *Igloi, *Cincio2007, *latorre, *1742-5468-2012-07-P07016}).
In terms of $\Pi$ the entanglement entropy of the sub-block of size $l$, can be  written  as 
$S_l(t)=-\operatorname{Tr}{(\Pi(t) \log_2 \Pi(t))}$.  

The numerical evaluation of  Floquet dynamics has been performed with the software QuTiP 3.1.0: The Quantum Toolbox in Python \cite{Nori}, while  the entanglement entropy  has been evaluated by numerical diagonalization of the $\Pi(t)$ matrix.

Having set  the mathematical framework we now describe our results for the dynamics and scaling of entanglement entropy, $S_l(t)$, after $n$ cycles of the driving. 
As we have followed the dynamics stroboscopically using Floquet formalism, 
the time dependence of entanglement entropy is measured in units of the driving period, $T$ and we  use the notation $S_l(n)$ to refer to the entanglement entropy after $n$ driving cycles.

After a short-time linear increase, the entanglement entropy, in analogy with observables in periodically driven integrable systems \cite{Angelo, Laz},  approaches a stationary value synchronised with the driving frequency $\omega$ \footnote{As a matter of fact, the only exception is represented by drivings starting at $h(t=0)=0$; in this case  entanglement entropy oscillates with  frequency $2\omega$.}. Furthermore, as a finite-size effect, we notice the emergence of quasi-revivals which break the synchronized plateaux.
%
%Furthermore, we found, by numerical investigation, that the maximal speed $v_{\max}$, for low frequency $\omega=\frac{\pi}{10}$, is a monotonically decreasing function both with the initial magnetic field $h(0)$ and the driving amplitude $\Delta h$, provided the latter is not too large, $\Delta h \simeq 1$, beyond which  $v_{\max}$ begins to fluctuate.
Similar revivals and short-time linear increase characterise also the entanglement entropy dynamics  induced by a sudden quantum quench of a QIC (see, for instance, Refs. \cite{Igloi, *igloi2}). Such dynamical features (short-time  increase, quasi-revivals,  and synchronisation), are robust to changes of the frequency or amplitude of the driving and they persist both when $h(t)$ is confined within the paramagnetic or ferromagnetic phase as well as when it crosses repeatedly these two phases.
% when $h(t)$ is centred at the critical point.
As an instance of this universal phenomenology, we reported the associated stroboscopic time evolution of the entanglement entropy for a driving inside the ferromagnetic phase and around the critical point $h=1$ in Fig. \ref{FigFM}. 

Interestingly, once the analogy between the Floquet modes and the quasi-particle of the final Hamiltonian in a sudden-quench scenario is set, we can adopt the picture of Floquet quasi-particles (FQPs) originating at each site, and a number of useful results can be found by simple cinematic considerations. 
The maximum propagation velocity of Floquet quasi-particles can be derived as follows. In contrast with the sudden-quench scenario~\cite{1742-5468-2012-07-P07016, *igloi2}, where the dispersion relation is given by the final Hamiltonian energy spectrum, for periodically driven systems the ''dispersion relation'' is given by the Floquet spectrum, which, as can be seen from Eq.~(\ref{E.Floquetstrobo}), rules the time evolution, yielding $v_{\max}=\underset{q\in[0,\pi]}{\max} \left|\frac{d\mu(q)}{dq}\right|$. 
%Instances of $v_{\max}$ are reported in Fig.~\ref{speed}.  
It follows that the duration $t_L$ of the linear short-time increase is determined by the Floquet quasi-particles leaving out the subsystem $l$, and hence is given by $t_L=l/v_{max}$. Clearly, the larger the subsystem size $l$ the longer the intermediate regime between the linear and the stationary regime, as slower quasi particles require more time to leave the region $l$. 
The recurrence time $t_R=(N-l)/(2v_{max}) $ is an accurate estimate of the time at which we observe the onset of the quasi-revivals, which are observed once the FQPs, originating at the opposite side with respect to the subsystem of size $l$ (hence at distance $(N-l)/2$), reach the subsystem. Now we can relate the different time scales appearing in Fig.~\ref{FigFM} to the spread of these quasi-particles for different drivings:
Corresponding to the frequency $\omega=\pi$, we see that the drive around $h=0$ and $h=1$ have, respectively, the lowest and fastest $v_{\max}$ (at fixed $\Delta h=0.5$), and this accounts for the longer (shorter) duration of the linear increase and one order of magnitude difference in the occurrence of the quasi-revivals effects  (cfr. with Fig. \ref{FigFM}).
Furthermore, we notice that $v_{max}$ is limited from above, $v_{max}<1$, signature of a bound of the quasiparticles propagation speed.

Let us now discuss the scaling of the entanglement entropy with the sub-system size, $l$, at long times i.e. in the plateau region (see Fig. \ref{FigFM}).

\begin{figure}[t!]\centering
\includegraphics[width=8.7cm]
{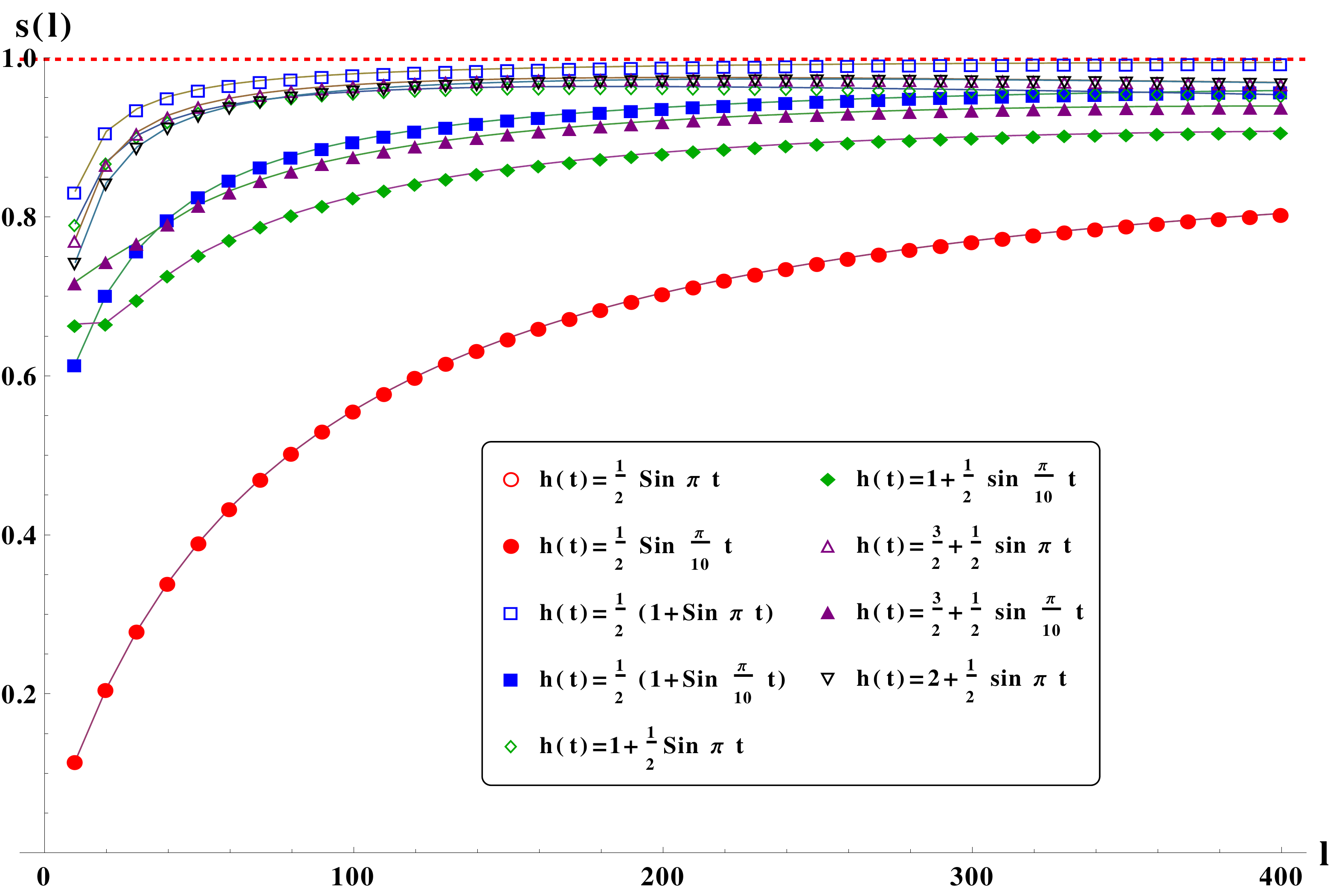}%{esse.pdf}
 \caption{(color online) Exponent $s\equiv \log[\bar{S}_{2l}/\bar{S}_l]/\log 2$ vs subsystem size $l$ ($N=8192$), obtained averaging over the stroboscopic entanglement entropy $S_l(nT)$ with $1\ll n < \frac{t_R}{T}$.  Different lines correspond to different realisations of the driving $h(t)$ (lines are for guiding the eyes).  
% 
%  \jm{commento sulla convergenza delle curve sfigate in caption; nota sul fatto che non vediamo un plateaux nel caso di bassa freq dentro il paramgnetico}
}
\label{esse}
\end{figure}

In order to extract the scaling of the entanglement entropy, with the sub-system size $l$,  we considered its stroboscopic time average $\bar{S}$, over the plateau associated to each single $l$,  and we found numerical evidence of the convergence of $S_l$ towards a volume law as the block size $l$ is increased. We quantified this asymptotic approach extracting the scaling form, $S_l\sim l^s$, controlled by the exponent $s\equiv \log[\bar{S}_{2l}/\bar{S}_l]/\log 2$, where $s=0$ ($s=1$) correspond to a area (volume) law scaling. %The stroboscopic average of the entanglement entropy (\jm{dire che indica in generale un rilassamento, non so se vada fatto visto che S non e' unpsservabile}) $\bar{\langle S_l(n)\rangle}=\lim_{n\to\infty}\frac{1}{n}\sum_{m=0}^{n-1}\langle S_l\rangle(t=mT)$, which washes out in the thermodynamics limit stroboscopic time-fluctuations and extract the synchronized long-time plateau of $S_n(l)$ in Fig.\ref{FigFM}. 
We numerically evaluated the approach towards a volume law for several realisations of the driving, $h(t)$. As clearly shown in Fig. \ref{esse}, $s(l)\rightarrow 1$ more rapidly for faster drivings. In the thermodynamic limit $l,N\rightarrow \infty$ and $l\ll N$, the numerical plots suggest a full establishment of the volume law \footnote{The driving  $h(t)=2+\frac{1}{2}\sin\frac{\pi}{10}t$ has not been included in Fig. \ref{esse} as $S_l(nT)$ exhibits a relaxation time beyond our numerical investigation and, thus, no plateaux appears in the dynamics of $S_l(t)$}. We investigated this conjecture numerically for the drivings exhibiting the slower approaches to $s(l)\rightarrow 1$, by increasing the system's size up to $N=4\,\text{x}\,10^4$ and subsystem's size up to $l=2\,\text{x}\,10^4$, confirming a monotonically increase  up to $s(l)\simeq 0.95$.

The Floquet quasi particle picture allows us to analyse the dynamics of $s_l(t)$ and provides a clear insight into the number of cycles $n$ at which the area law is violated. We identify three ($l$-dependent) time intervals. In the first, $t\in [0, \frac{l}{v_{max}}]$, $s_l(t)=0$ because both $S_l(t)$ and $S_{2l}(t)$ increase linearly in time by the same amount; at $t\simeq\frac{l}{v_{max}}$, $S_l(t)$ has reached its stationary value, while $S_{2l}(t)$ continues its linear increase until $t\simeq\frac{2l}{v_{max}}$, resulting in a logarithmic increase of $s_l(t)$, which stops at $t\simeq t_R$, when also  $S_{2l}(t)$ stabilises and $s_l(t)$ attains the value reported in Fig.~\ref{esse}. 
As a consequence, the breakdown of the area law (or of the logarithmic scaling, if we start at criticality~\cite{IgloiLinJSM08}) occurs the faster the larger the values of $v_{max}$.
For the settings we are here considering, at $\omega=\frac{\pi}{10}$ violation occurs faster by increasing the initial magnetic field $h(0)$, whereas, for the higher frequency considered in our analysis, $\omega=\pi$, it is faster at criticality.

\begin{figure}[t!]\centering
\includegraphics[width=\linewidth]
{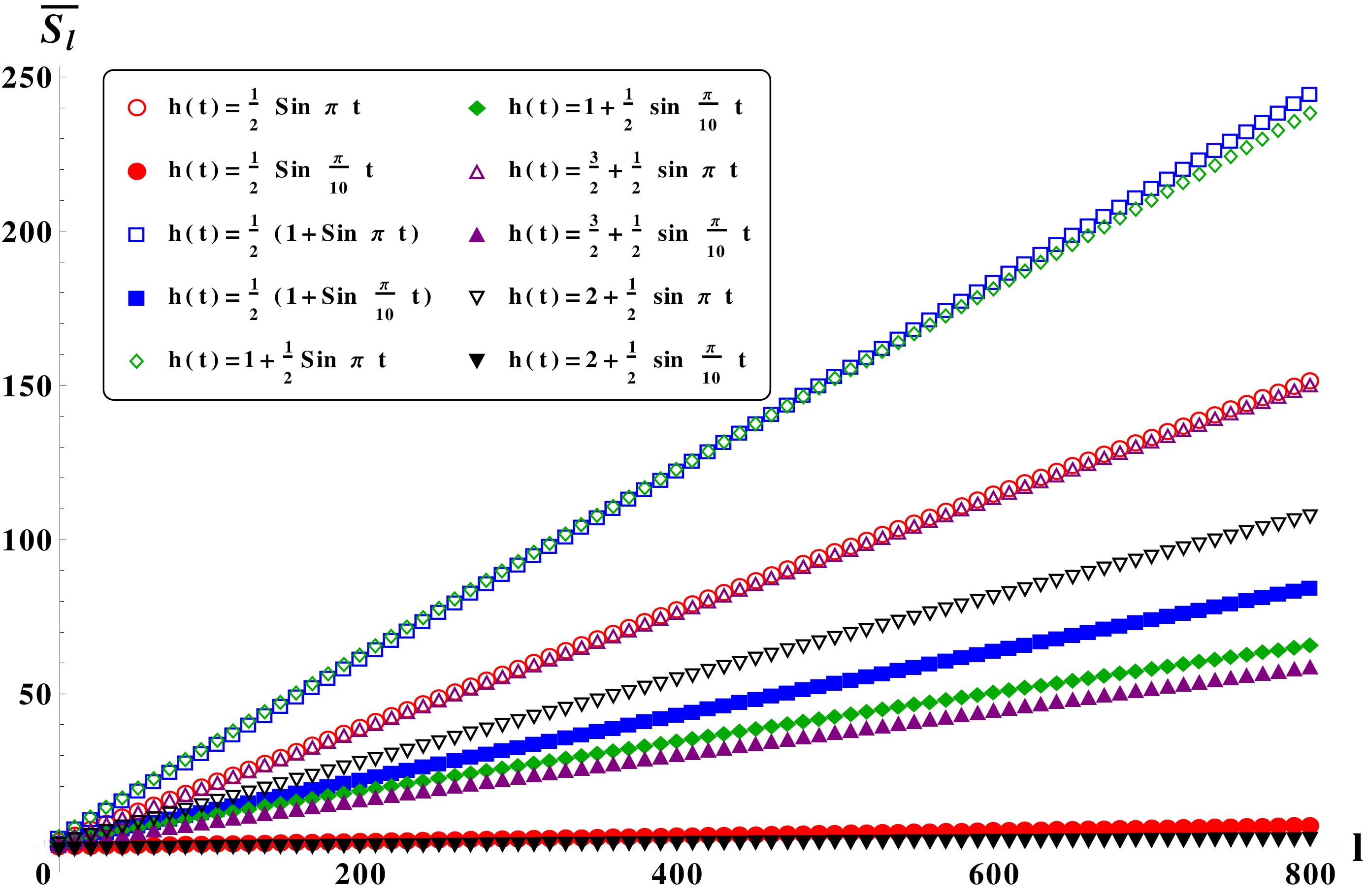}%{esse.pdf}
 \caption{(color online) Asymptotic time-averaged entanglement entropy vs subsystem size $l$ for different driving protocols. 
}
\label{maxEE}
\end{figure}

Finally, considering the asymptotic value of entanglement entropy vs $l$, reported in Fig.~\ref{maxEE} for a spin ring of $N=8192$ spins, we notice that it remains well below the infinite temperature value, where $S_l=l$, showing indirectly that the system does not heat up indefinitely. 
In addition, let us also point out that neither the amount of asymptotic entanglement entropy and the scaling exponents $s$  (see Fig.~\ref{esse}) appear to be correlated, as different drivings $h(t)$ may yield the same value of $\bar{S}_l$, despite the fact that the corresponding $s$ are different (or viceversa); nor higher values of $s$ correspond to higher values of $\bar{S}_l$. This means that the proportionality constant in $\bar{S}_l \sim l^s$ should be protocol-dependent.

%\changer{We notice also that faster drivings yield an higher amount of entanglement entropy with respect to slower ones and that comparable values of the exponent $s_l$ in Fig.~\ref{esse} do not correspond to comparable values of $S_l^{max}$, meaning that the coefficient in $S_l \sim l^s$ is protocol-dependent.}

In conclusion, we have shown that the asymptotic entanglement entropy scales in accordance with a volume law for a broad range of frequency, amplitude, and average value of the periodically driven transverse field$h(t)$. This result strengthens the analogy between  periodically driven and suddenly quenched  quantum many-particle systems, and it calls for similar studies in other classes of periodically driven systems which are not expected to undergo to heating leading to an infinite temperature state. For instance, it would be interesting to check whether similar results hold for integrable models not equivalent to quadratic fermions/bosons, or for interacting systems in the presence of spatial disorder \cite{CC, Roy, LazMBL}, which can host novel phases with the potential for exotic scaling of entanglement entropy \cite{Sondhi}.

While this manuscript was in preparation, we became aware of similar results obtained by A. Russomanno, G.~E. Santoro, and R. Fazio \cite{AngeloNEW}.

\emph{Acknowledgments.} 
%We  thank  A. Russomanno for providing numerical data from Ref. \cite{Angelo}, which have been used as a benchmark for our Python code. 
We thank  M. Fagotti, R. Fazio, M. Genske, A. Lazarides, M. Marcuzzi, A. Russomanno, E. Tonni, for useful discussions, A. Silva for inspiring insights, and S. Lorenzo for crucial help with the Floquet analysis. J. M.  acknowledges financial support from the Alexander Von Humboldt Foundation. T.J.G.A. and G.M.P acknowledge funding  under the EU Collaborative Project TherMiQ (Grant No. 618074) and the EU grant  QuPRoCs (Grant Agreement 641277) .

\bibliography{Drivenbiblio.bib}

\begin{thebibliography}{82}%
\makeatletter
\providecommand \@ifxundefined [1]{%
 \@ifx{#1\undefined}
}%
\providecommand \@ifnum [1]{%
 \ifnum #1\expandafter \@firstoftwo
 \else \expandafter \@secondoftwo
 \fi
}%
\providecommand \@ifx [1]{%
 \ifx #1\expandafter \@firstoftwo
 \else \expandafter \@secondoftwo
 \fi
}%
\providecommand \natexlab [1]{#1}%
\providecommand \enquote  [1]{``#1''}%
\providecommand \bibnamefont  [1]{#1}%
\providecommand \bibfnamefont [1]{#1}%
\providecommand \citenamefont [1]{#1}%
\providecommand \href@noop [0]{\@secondoftwo}%
\providecommand \href [0]{\begingroup \@sanitize@url \@href}%
\providecommand \@href[1]{\@@startlink{#1}\@@href}%
\providecommand \@@href[1]{\endgroup#1\@@endlink}%
\providecommand \@sanitize@url [0]{\catcode `\\12\catcode `\$12\catcode
  `\&12\catcode `\#12\catcode `\^12\catcode `\_12\catcode `\%12\relax}%
\providecommand \@@startlink[1]{}%
\providecommand \@@endlink[0]{}%
\providecommand \url  [0]{\begingroup\@sanitize@url \@url }%
\providecommand \@url [1]{\endgroup\@href {#1}{\urlprefix }}%
\providecommand \urlprefix  [0]{URL }%
\providecommand \Eprint [0]{\href }%
\providecommand \doibase [0]{http://dx.doi.org/}%
\providecommand \selectlanguage [0]{\@gobble}%
\providecommand \bibinfo  [0]{\@secondoftwo}%
\providecommand \bibfield  [0]{\@secondoftwo}%
\providecommand \translation [1]{[#1]}%
\providecommand \BibitemOpen [0]{}%
\providecommand \bibitemStop [0]{}%
\providecommand \bibitemNoStop [0]{.\EOS\space}%
\providecommand \EOS [0]{\spacefactor3000\relax}%
\providecommand \BibitemShut  [1]{\csname bibitem#1\endcsname}%
\let\auto@bib@innerbib\@empty
%</preamble>
\bibitem [{\citenamefont {Bloch}\ \emph {et~al.}(2008)\citenamefont {Bloch},
  \citenamefont {Dalibard},\ and\ \citenamefont {Zwerger}}]{Zwerger}%
  \BibitemOpen
  \bibfield  {author} {\bibinfo {author} {\bibfnamefont {I.}~\bibnamefont
  {Bloch}}, \bibinfo {author} {\bibfnamefont {J.}~\bibnamefont {Dalibard}}, \
  and\ \bibinfo {author} {\bibfnamefont {W.}~\bibnamefont {Zwerger}},\ }\href
  {\doibase 10.1103/RevModPhys.80.885} {\bibfield  {journal} {\bibinfo
  {journal} {Rev. Mod. Phys.}\ }\textbf {\bibinfo {volume} {80}},\ \bibinfo
  {pages} {885} (\bibinfo {year} {2008})}\BibitemShut {NoStop}%
\bibitem [{\citenamefont {Polkovnikov}\ \emph {et~al.}(2011)\citenamefont
  {Polkovnikov}, \citenamefont {Sengupta}, \citenamefont {Silva},\ and\
  \citenamefont {Vengalattore}}]{Silva}%
  \BibitemOpen
  \bibfield  {author} {\bibinfo {author} {\bibfnamefont {A.}~\bibnamefont
  {Polkovnikov}}, \bibinfo {author} {\bibfnamefont {K.}~\bibnamefont
  {Sengupta}}, \bibinfo {author} {\bibfnamefont {A.}~\bibnamefont {Silva}}, \
  and\ \bibinfo {author} {\bibfnamefont {M.}~\bibnamefont {Vengalattore}},\
  }\href {\doibase 10.1103/RevModPhys.83.863} {\bibfield  {journal} {\bibinfo
  {journal} {Rev. Mod. Phys.}\ }\textbf {\bibinfo {volume} {83}},\ \bibinfo
  {pages} {863} (\bibinfo {year} {2011})}\BibitemShut {NoStop}%
\bibitem [{\citenamefont {Gogolin}\ and\ \citenamefont
  {Eisert}(2015)}]{Eisert}%
  \BibitemOpen
  \bibfield  {author} {\bibinfo {author} {\bibfnamefont {C.}~\bibnamefont
  {Gogolin}}\ and\ \bibinfo {author} {\bibfnamefont {J.}~\bibnamefont
  {Eisert}},\ }\href@noop {} {\bibfield  {journal} {\bibinfo  {journal}
  {arXiv:1503.07538}\ } (\bibinfo {year} {2015})}\BibitemShut {NoStop}%
\bibitem [{\citenamefont {Sieberer}\ \emph {et~al.}(2015)\citenamefont
  {Sieberer}, \citenamefont {Buchhold},\ and\ \citenamefont {Diehl}}]{Diehl}%
  \BibitemOpen
  \bibfield  {author} {\bibinfo {author} {\bibfnamefont {L.~M.}\ \bibnamefont
  {Sieberer}}, \bibinfo {author} {\bibfnamefont {M.}~\bibnamefont {Buchhold}},
  \ and\ \bibinfo {author} {\bibfnamefont {S.}~\bibnamefont {Diehl}},\
  }\href@noop {} {\bibfield  {journal} {\bibinfo  {journal} {arXiv:1512.00637}\
  } (\bibinfo {year} {2015})}\BibitemShut {NoStop}%
\bibitem [{\citenamefont {Deutsch}(1991)}]{Deutsch}%
  \BibitemOpen
  \bibfield  {author} {\bibinfo {author} {\bibfnamefont {J.~M.}\ \bibnamefont
  {Deutsch}},\ }\href {\doibase 10.1103/PhysRevA.43.2046} {\bibfield  {journal}
  {\bibinfo  {journal} {Phys. Rev. A}\ }\textbf {\bibinfo {volume} {43}},\
  \bibinfo {pages} {2046} (\bibinfo {year} {1991})}\BibitemShut {NoStop}%
\bibitem [{\citenamefont {Srednicki}(1994)}]{Srednicki}%
  \BibitemOpen
  \bibfield  {author} {\bibinfo {author} {\bibfnamefont {M.}~\bibnamefont
  {Srednicki}},\ }\href {\doibase 10.1103/PhysRevE.50.888} {\bibfield
  {journal} {\bibinfo  {journal} {Phys. Rev. E}\ }\textbf {\bibinfo {volume}
  {50}},\ \bibinfo {pages} {888} (\bibinfo {year} {1994})}\BibitemShut
  {NoStop}%
\bibitem [{\citenamefont {Rigol}\ \emph {et~al.}(2008)\citenamefont {Rigol},
  \citenamefont {Dunjko},\ and\ \citenamefont {Olshanii}}]{Rigol}%
  \BibitemOpen
  \bibfield  {author} {\bibinfo {author} {\bibfnamefont {M.}~\bibnamefont
  {Rigol}}, \bibinfo {author} {\bibfnamefont {V.}~\bibnamefont {Dunjko}}, \
  and\ \bibinfo {author} {\bibfnamefont {M.}~\bibnamefont {Olshanii}},\
  }\href@noop {} {\bibfield  {journal} {\bibinfo  {journal} {Nature}\ }\textbf
  {\bibinfo {volume} {452}},\ \bibinfo {pages} {854} (\bibinfo {year}
  {2008})}\BibitemShut {NoStop}%
\bibitem [{\citenamefont {Grifoni}\ and\ \citenamefont
  {H\"anggi}(1998)}]{Grifoni}%
  \BibitemOpen
  \bibfield  {author} {\bibinfo {author} {\bibfnamefont {M.~G.}\ \bibnamefont
  {Grifoni}}\ and\ \bibinfo {author} {\bibfnamefont {P.}~\bibnamefont
  {H\"anggi}},\ }\href@noop {} {\bibfield  {journal} {\bibinfo  {journal}
  {Phys. Rep.}\ }\textbf {\bibinfo {volume} {304}},\ \bibinfo {pages} {229}
  (\bibinfo {year} {1998})}\BibitemShut {NoStop}%
\bibitem [{\citenamefont {Eckardt}\ \emph {et~al.}(2005)\citenamefont
  {Eckardt}, \citenamefont {Weiss},\ and\ \citenamefont {Holthaus}}]{Weiss05}%
  \BibitemOpen
  \bibfield  {author} {\bibinfo {author} {\bibfnamefont {A.}~\bibnamefont
  {Eckardt}}, \bibinfo {author} {\bibfnamefont {C.}~\bibnamefont {Weiss}}, \
  and\ \bibinfo {author} {\bibfnamefont {M.}~\bibnamefont {Holthaus}},\ }\href
  {\doibase 10.1103/PhysRevLett.95.260404} {\bibfield  {journal} {\bibinfo
  {journal} {Phys. Rev. Lett.}\ }\textbf {\bibinfo {volume} {95}},\ \bibinfo
  {pages} {260404} (\bibinfo {year} {2005})}\BibitemShut {NoStop}%
\bibitem [{\citenamefont {Sias}\ \emph {et~al.}(2007)\citenamefont {Sias},
  \citenamefont {Zenesini}, \citenamefont {Lignier}, \citenamefont {Wimberger},
  \citenamefont {Ciampini}, \citenamefont {Morsch},\ and\ \citenamefont
  {Arimondo}}]{Sias07}%
  \BibitemOpen
  \bibfield  {author} {\bibinfo {author} {\bibfnamefont {C.}~\bibnamefont
  {Sias}}, \bibinfo {author} {\bibfnamefont {A.}~\bibnamefont {Zenesini}},
  \bibinfo {author} {\bibfnamefont {H.}~\bibnamefont {Lignier}}, \bibinfo
  {author} {\bibfnamefont {S.}~\bibnamefont {Wimberger}}, \bibinfo {author}
  {\bibfnamefont {D.}~\bibnamefont {Ciampini}}, \bibinfo {author}
  {\bibfnamefont {O.}~\bibnamefont {Morsch}}, \ and\ \bibinfo {author}
  {\bibfnamefont {E.}~\bibnamefont {Arimondo}},\ }\href {\doibase
  10.1103/PhysRevLett.98.120403} {\bibfield  {journal} {\bibinfo  {journal}
  {Phys. Rev. Lett.}\ }\textbf {\bibinfo {volume} {98}},\ \bibinfo {pages}
  {120403} (\bibinfo {year} {2007})}\BibitemShut {NoStop}%
\bibitem [{\citenamefont {Zenesini}\ \emph {et~al.}(2009)\citenamefont
  {Zenesini}, \citenamefont {Lignier}, \citenamefont {Ciampini}, \citenamefont
  {Morsch},\ and\ \citenamefont {Arimondo}}]{Zenesini09}%
  \BibitemOpen
  \bibfield  {author} {\bibinfo {author} {\bibfnamefont {A.}~\bibnamefont
  {Zenesini}}, \bibinfo {author} {\bibfnamefont {H.}~\bibnamefont {Lignier}},
  \bibinfo {author} {\bibfnamefont {D.}~\bibnamefont {Ciampini}}, \bibinfo
  {author} {\bibfnamefont {O.}~\bibnamefont {Morsch}}, \ and\ \bibinfo {author}
  {\bibfnamefont {E.}~\bibnamefont {Arimondo}},\ }\href {\doibase
  10.1103/PhysRevLett.102.100403} {\bibfield  {journal} {\bibinfo  {journal}
  {Phys. Rev. Lett.}\ }\textbf {\bibinfo {volume} {102}},\ \bibinfo {pages}
  {100403} (\bibinfo {year} {2009})}\BibitemShut {NoStop}%
\bibitem [{\citenamefont {Struck}\ \emph {et~al.}(2011)\citenamefont {Struck},
  \citenamefont {Ölschläger}, \citenamefont {Le~Targat}, \citenamefont
  {Soltan-Panahi}, \citenamefont {Eckardt}, \citenamefont {Lewenstein},
  \citenamefont {Windpassinger},\ and\ \citenamefont {K.}}]{Struck11}%
  \BibitemOpen
  \bibfield  {author} {\bibinfo {author} {\bibfnamefont {J.}~\bibnamefont
  {Struck}}, \bibinfo {author} {\bibfnamefont {C.}~\bibnamefont {Ölschläger}},
  \bibinfo {author} {\bibfnamefont {R.}~\bibnamefont {Le~Targat}}, \bibinfo
  {author} {\bibfnamefont {P.}~\bibnamefont {Soltan-Panahi}}, \bibinfo {author}
  {\bibfnamefont {A.}~\bibnamefont {Eckardt}}, \bibinfo {author} {\bibfnamefont
  {M.}~\bibnamefont {Lewenstein}}, \bibinfo {author} {\bibfnamefont
  {P.}~\bibnamefont {Windpassinger}}, \ and\ \bibinfo {author} {\bibfnamefont
  {S.}~\bibnamefont {K.}},\ }\href@noop {} {\bibfield  {journal} {\bibinfo
  {journal} {Science}\ }\textbf {\bibinfo {volume} {333}},\ \bibinfo {pages}
  {996} (\bibinfo {year} {2011})}\BibitemShut {NoStop}%
\bibitem [{\citenamefont {Miyake}\ \emph {et~al.}(2013)\citenamefont {Miyake},
  \citenamefont {Siviloglou}, \citenamefont {Kennedy}, \citenamefont {Burton},\
  and\ \citenamefont {Ketterle}}]{Burton13}%
  \BibitemOpen
  \bibfield  {author} {\bibinfo {author} {\bibfnamefont {H.}~\bibnamefont
  {Miyake}}, \bibinfo {author} {\bibfnamefont {G.~A.}\ \bibnamefont
  {Siviloglou}}, \bibinfo {author} {\bibfnamefont {C.~J.}\ \bibnamefont
  {Kennedy}}, \bibinfo {author} {\bibfnamefont {W.~C.}\ \bibnamefont {Burton}},
  \ and\ \bibinfo {author} {\bibfnamefont {W.}~\bibnamefont {Ketterle}},\
  }\href {\doibase 10.1103/PhysRevLett.111.185302} {\bibfield  {journal}
  {\bibinfo  {journal} {Phys. Rev. Lett.}\ }\textbf {\bibinfo {volume} {111}},\
  \bibinfo {pages} {185302} (\bibinfo {year} {2013})}\BibitemShut {NoStop}%
\bibitem [{\citenamefont {Jacksh}\ and\ \citenamefont
  {Zoller}(2003)}]{Jacksh03}%
  \BibitemOpen
  \bibfield  {author} {\bibinfo {author} {\bibfnamefont {D.}~\bibnamefont
  {Jacksh}}\ and\ \bibinfo {author} {\bibfnamefont {P.}~\bibnamefont
  {Zoller}},\ }\href@noop {} {\bibfield  {journal} {\bibinfo  {journal} {Ann.
  Phys.}\ }\textbf {\bibinfo {volume} {315}},\ \bibinfo {pages} {52} (\bibinfo
  {year} {2003})}\BibitemShut {NoStop}%
\bibitem [{\citenamefont {Struck}\ \emph {et~al.}(2013)\citenamefont {Struck},
  \citenamefont {Weinberg}, \citenamefont {Olschlager}, \citenamefont
  {Windpassinger}, \citenamefont {Simonet}, \citenamefont {Sengstock},
  \citenamefont {Hoppner}, \citenamefont {Hauke}, \citenamefont {Eckardt},
  \citenamefont {Lewenstein},\ and\ \citenamefont {Mathey}}]{Struck13}%
  \BibitemOpen
  \bibfield  {author} {\bibinfo {author} {\bibfnamefont {J.}~\bibnamefont
  {Struck}}, \bibinfo {author} {\bibfnamefont {M.}~\bibnamefont {Weinberg}},
  \bibinfo {author} {\bibfnamefont {C.}~\bibnamefont {Olschlager}}, \bibinfo
  {author} {\bibfnamefont {P.}~\bibnamefont {Windpassinger}}, \bibinfo {author}
  {\bibfnamefont {J.}~\bibnamefont {Simonet}}, \bibinfo {author} {\bibfnamefont
  {K.}~\bibnamefont {Sengstock}}, \bibinfo {author} {\bibfnamefont
  {R.}~\bibnamefont {Hoppner}}, \bibinfo {author} {\bibfnamefont
  {P.}~\bibnamefont {Hauke}}, \bibinfo {author} {\bibfnamefont
  {A.}~\bibnamefont {Eckardt}}, \bibinfo {author} {\bibfnamefont
  {M.}~\bibnamefont {Lewenstein}}, \ and\ \bibinfo {author} {\bibfnamefont
  {L.}~\bibnamefont {Mathey}},\ }\href {http://dx.doi.org/10.1038/nphys2750}
  {\bibfield  {journal} {\bibinfo  {journal} {Nat. Phys.}\ }\textbf {\bibinfo
  {volume} {9}},\ \bibinfo {pages} {738} (\bibinfo {year} {2013})}\BibitemShut
  {NoStop}%
\bibitem [{\citenamefont {Aidelsburger}\ \emph {et~al.}(2013)\citenamefont
  {Aidelsburger}, \citenamefont {Atala}, \citenamefont {Lohse}, \citenamefont
  {Barreiro}, \citenamefont {Paredes},\ and\ \citenamefont {Bloch}}]{Atala13}%
  \BibitemOpen
  \bibfield  {author} {\bibinfo {author} {\bibfnamefont {M.}~\bibnamefont
  {Aidelsburger}}, \bibinfo {author} {\bibfnamefont {M.}~\bibnamefont {Atala}},
  \bibinfo {author} {\bibfnamefont {M.}~\bibnamefont {Lohse}}, \bibinfo
  {author} {\bibfnamefont {J.~T.}\ \bibnamefont {Barreiro}}, \bibinfo {author}
  {\bibfnamefont {B.}~\bibnamefont {Paredes}}, \ and\ \bibinfo {author}
  {\bibfnamefont {I.}~\bibnamefont {Bloch}},\ }\href {\doibase
  10.1103/PhysRevLett.111.185301} {\bibfield  {journal} {\bibinfo  {journal}
  {Phys. Rev. Lett.}\ }\textbf {\bibinfo {volume} {111}},\ \bibinfo {pages}
  {185301} (\bibinfo {year} {2013})}\BibitemShut {NoStop}%
\bibitem [{\citenamefont {Goldman}\ and\ \citenamefont
  {Dalibard}(2014)}]{Goldman14}%
  \BibitemOpen
  \bibfield  {author} {\bibinfo {author} {\bibfnamefont {N.}~\bibnamefont
  {Goldman}}\ and\ \bibinfo {author} {\bibfnamefont {J.}~\bibnamefont
  {Dalibard}},\ }\href {\doibase 10.1103/PhysRevX.4.031027} {\bibfield
  {journal} {\bibinfo  {journal} {Phys. Rev. X}\ }\textbf {\bibinfo {volume}
  {4}},\ \bibinfo {pages} {031027} (\bibinfo {year} {2014})}\BibitemShut
  {NoStop}%
\bibitem [{\citenamefont {Oka}\ and\ \citenamefont {Aoki}(2009)}]{Oka09}%
  \BibitemOpen
  \bibfield  {author} {\bibinfo {author} {\bibfnamefont {T.}~\bibnamefont
  {Oka}}\ and\ \bibinfo {author} {\bibfnamefont {H.}~\bibnamefont {Aoki}},\
  }\href {\doibase 10.1103/PhysRevB.79.081406} {\bibfield  {journal} {\bibinfo
  {journal} {Phys. Rev. B}\ }\textbf {\bibinfo {volume} {79}},\ \bibinfo
  {pages} {081406} (\bibinfo {year} {2009})}\BibitemShut {NoStop}%
\bibitem [{\citenamefont {Kitagawa}\ \emph {et~al.}(2011)\citenamefont
  {Kitagawa}, \citenamefont {Oka}, \citenamefont {Brataas}, \citenamefont
  {Fu},\ and\ \citenamefont {Demler}}]{Kitagawa11}%
  \BibitemOpen
  \bibfield  {author} {\bibinfo {author} {\bibfnamefont {T.}~\bibnamefont
  {Kitagawa}}, \bibinfo {author} {\bibfnamefont {T.}~\bibnamefont {Oka}},
  \bibinfo {author} {\bibfnamefont {A.}~\bibnamefont {Brataas}}, \bibinfo
  {author} {\bibfnamefont {L.}~\bibnamefont {Fu}}, \ and\ \bibinfo {author}
  {\bibfnamefont {E.}~\bibnamefont {Demler}},\ }\href {\doibase
  10.1103/PhysRevB.84.235108} {\bibfield  {journal} {\bibinfo  {journal} {Phys.
  Rev. B}\ }\textbf {\bibinfo {volume} {84}},\ \bibinfo {pages} {235108}
  (\bibinfo {year} {2011})}\BibitemShut {NoStop}%
\bibitem [{\citenamefont {Jotzu}\ \emph {et~al.}(2014)\citenamefont {Jotzu},
  \citenamefont {Messer}, \citenamefont {Desbuquois}, \citenamefont {Lebrat},
  \citenamefont {Uehlinger}, \citenamefont {Greif},\ and\ \citenamefont
  {Esslinger}}]{Jotzu14}%
  \BibitemOpen
  \bibfield  {author} {\bibinfo {author} {\bibfnamefont {G.}~\bibnamefont
  {Jotzu}}, \bibinfo {author} {\bibfnamefont {M.}~\bibnamefont {Messer}},
  \bibinfo {author} {\bibfnamefont {R.}~\bibnamefont {Desbuquois}}, \bibinfo
  {author} {\bibfnamefont {M.}~\bibnamefont {Lebrat}}, \bibinfo {author}
  {\bibfnamefont {T.}~\bibnamefont {Uehlinger}}, \bibinfo {author}
  {\bibfnamefont {D.}~\bibnamefont {Greif}}, \ and\ \bibinfo {author}
  {\bibfnamefont {T.}~\bibnamefont {Esslinger}},\ }\href@noop {} {\bibfield
  {journal} {\bibinfo  {journal} {Nature}\ }\textbf {\bibinfo {volume} {515}},\
  \bibinfo {pages} {237} (\bibinfo {year} {2014})}\BibitemShut {NoStop}%
\bibitem [{\citenamefont {Aidelsburger}\ \emph {et~al.}(2015)\citenamefont
  {Aidelsburger}, \citenamefont {Lohse}, \citenamefont {Schweizer},
  \citenamefont {Atala}, \citenamefont {Barreiro}, \citenamefont {Nascimbene},
  \citenamefont {Cooper}, \citenamefont {Bloch},\ and\ \citenamefont
  {Goldman}}]{Bloch15}%
  \BibitemOpen
  \bibfield  {author} {\bibinfo {author} {\bibfnamefont {M.}~\bibnamefont
  {Aidelsburger}}, \bibinfo {author} {\bibfnamefont {M.}~\bibnamefont {Lohse}},
  \bibinfo {author} {\bibfnamefont {C.}~\bibnamefont {Schweizer}}, \bibinfo
  {author} {\bibfnamefont {M.}~\bibnamefont {Atala}}, \bibinfo {author}
  {\bibfnamefont {J.~T.}\ \bibnamefont {Barreiro}}, \bibinfo {author}
  {\bibfnamefont {S.}~\bibnamefont {Nascimbene}}, \bibinfo {author}
  {\bibfnamefont {N.~R.}\ \bibnamefont {Cooper}}, \bibinfo {author}
  {\bibfnamefont {I.}~\bibnamefont {Bloch}}, \ and\ \bibinfo {author}
  {\bibfnamefont {N.}~\bibnamefont {Goldman}},\ }\href@noop {} {\bibfield
  {journal} {\bibinfo  {journal} {Nature Physics}\ }\textbf {\bibinfo {volume}
  {11}},\ \bibinfo {pages} {162} (\bibinfo {year} {2015})}\BibitemShut
  {NoStop}%
\bibitem [{\citenamefont {Lindner}\ \emph {et~al.}(2011)\citenamefont
  {Lindner}, \citenamefont {Refael},\ and\ \citenamefont
  {Galitski}}]{Refael11}%
  \BibitemOpen
  \bibfield  {author} {\bibinfo {author} {\bibfnamefont {N.~H.}\ \bibnamefont
  {Lindner}}, \bibinfo {author} {\bibfnamefont {G.}~\bibnamefont {Refael}}, \
  and\ \bibinfo {author} {\bibfnamefont {V.}~\bibnamefont {Galitski}},\
  }\href@noop {} {\bibfield  {journal} {\bibinfo  {journal} {Nature Physics}\
  }\textbf {\bibinfo {volume} {7}},\ \bibinfo {pages} {490} (\bibinfo {year}
  {2011})}\BibitemShut {NoStop}%
\bibitem [{\citenamefont {Rudner}\ \emph {et~al.}(2013)\citenamefont {Rudner},
  \citenamefont {Lindner}, \citenamefont {Berg},\ and\ \citenamefont
  {Levin}}]{Levin13}%
  \BibitemOpen
  \bibfield  {author} {\bibinfo {author} {\bibfnamefont {M.~S.}\ \bibnamefont
  {Rudner}}, \bibinfo {author} {\bibfnamefont {N.~H.}\ \bibnamefont {Lindner}},
  \bibinfo {author} {\bibfnamefont {E.}~\bibnamefont {Berg}}, \ and\ \bibinfo
  {author} {\bibfnamefont {M.}~\bibnamefont {Levin}},\ }\href {\doibase
  10.1103/PhysRevX.3.031005} {\bibfield  {journal} {\bibinfo  {journal} {Phys.
  Rev. X}\ }\textbf {\bibinfo {volume} {3}},\ \bibinfo {pages} {031005}
  (\bibinfo {year} {2013})}\BibitemShut {NoStop}%
\bibitem [{\citenamefont {Wang}\ \emph {et~al.}(2013)\citenamefont {Wang},
  \citenamefont {Steinberg}, \citenamefont {Jarillo-Herrero},\ and\
  \citenamefont {Gedik}}]{Wang13}%
  \BibitemOpen
  \bibfield  {author} {\bibinfo {author} {\bibfnamefont {Y.~H.}\ \bibnamefont
  {Wang}}, \bibinfo {author} {\bibfnamefont {H.}~\bibnamefont {Steinberg}},
  \bibinfo {author} {\bibfnamefont {P.}~\bibnamefont {Jarillo-Herrero}}, \ and\
  \bibinfo {author} {\bibfnamefont {N.}~\bibnamefont {Gedik}},\ }\href@noop {}
  {\bibfield  {journal} {\bibinfo  {journal} {Science}\ }\textbf {\bibinfo
  {volume} {342}},\ \bibinfo {pages} {453} (\bibinfo {year}
  {2013})}\BibitemShut {NoStop}%
\bibitem [{\citenamefont {Iadecola}\ \emph {et~al.}(2013)\citenamefont
  {Iadecola}, \citenamefont {Campbell}, \citenamefont {Chamon}, \citenamefont
  {Hou}, \citenamefont {Jackiw}, \citenamefont {Pi},\ and\ \citenamefont
  {Kusminskiy}}]{Iadecola15}%
  \BibitemOpen
  \bibfield  {author} {\bibinfo {author} {\bibfnamefont {T.}~\bibnamefont
  {Iadecola}}, \bibinfo {author} {\bibfnamefont {D.}~\bibnamefont {Campbell}},
  \bibinfo {author} {\bibfnamefont {C.}~\bibnamefont {Chamon}}, \bibinfo
  {author} {\bibfnamefont {C.-Y.}\ \bibnamefont {Hou}}, \bibinfo {author}
  {\bibfnamefont {R.}~\bibnamefont {Jackiw}}, \bibinfo {author} {\bibfnamefont
  {S.-Y.}\ \bibnamefont {Pi}}, \ and\ \bibinfo {author} {\bibfnamefont {S.~V.}\
  \bibnamefont {Kusminskiy}},\ }\href {\doibase 10.1103/PhysRevLett.110.176603}
  {\bibfield  {journal} {\bibinfo  {journal} {Phys. Rev. Lett.}\ }\textbf
  {\bibinfo {volume} {110}},\ \bibinfo {pages} {176603} (\bibinfo {year}
  {2013})}\BibitemShut {NoStop}%
\bibitem [{\citenamefont {D'Alessio}\ and\ \citenamefont
  {Rigol}(2015)}]{RigolChern}%
  \BibitemOpen
  \bibfield  {author} {\bibinfo {author} {\bibfnamefont {L.}~\bibnamefont
  {D'Alessio}}\ and\ \bibinfo {author} {\bibfnamefont {M.}~\bibnamefont
  {Rigol}},\ }\href@noop {} {\bibfield  {journal} {\bibinfo  {journal} {Nat.
  Commun.}\ }\textbf {\bibinfo {volume} {6}} (\bibinfo {year}
  {2015})}\BibitemShut {NoStop}%
\bibitem [{\citenamefont {D'Alessio}\ and\ \citenamefont
  {Polkovnikov}(2013)}]{Dal3}%
  \BibitemOpen
  \bibfield  {author} {\bibinfo {author} {\bibfnamefont {L.}~\bibnamefont
  {D'Alessio}}\ and\ \bibinfo {author} {\bibfnamefont {A.}~\bibnamefont
  {Polkovnikov}},\ }\href@noop {} {\bibfield  {journal} {\bibinfo  {journal}
  {Annals of Physics}\ }\textbf {\bibinfo {volume} {333}} (\bibinfo {year}
  {2013})}\BibitemShut {NoStop}%
\bibitem [{\citenamefont {D'Alessio}\ and\ \citenamefont
  {Rigol}(2014)}]{Rigol14}%
  \BibitemOpen
  \bibfield  {author} {\bibinfo {author} {\bibfnamefont {L.}~\bibnamefont
  {D'Alessio}}\ and\ \bibinfo {author} {\bibfnamefont {M.}~\bibnamefont
  {Rigol}},\ }\href {\doibase 10.1103/PhysRevX.4.041048} {\bibfield  {journal}
  {\bibinfo  {journal} {Phys. Rev. X}\ }\textbf {\bibinfo {volume} {4}},\
  \bibinfo {pages} {041048} (\bibinfo {year} {2014})}\BibitemShut {NoStop}%
\bibitem [{\citenamefont {Seetharam}\ \emph {et~al.}(2015)\citenamefont
  {Seetharam}, \citenamefont {Bardyn}, \citenamefont {Lindner}, \citenamefont
  {Rudner},\ and\ \citenamefont {Refael}}]{Refael}%
  \BibitemOpen
  \bibfield  {author} {\bibinfo {author} {\bibfnamefont {K.~I.}\ \bibnamefont
  {Seetharam}}, \bibinfo {author} {\bibfnamefont {C.-E.}\ \bibnamefont
  {Bardyn}}, \bibinfo {author} {\bibfnamefont {N.~H.}\ \bibnamefont {Lindner}},
  \bibinfo {author} {\bibfnamefont {M.~S.}\ \bibnamefont {Rudner}}, \ and\
  \bibinfo {author} {\bibfnamefont {G.}~\bibnamefont {Refael}},\ }\href
  {\doibase 10.1103/PhysRevX.5.041050} {\bibfield  {journal} {\bibinfo
  {journal} {Phys. Rev. X}\ }\textbf {\bibinfo {volume} {5}},\ \bibinfo {pages}
  {041050} (\bibinfo {year} {2015})}\BibitemShut {NoStop}%
\bibitem [{\citenamefont {Lazarides}\ \emph
  {et~al.}(2014{\natexlab{a}})\citenamefont {Lazarides}, \citenamefont {Das},\
  and\ \citenamefont {Moessner}}]{Lazarides14PRE}%
  \BibitemOpen
  \bibfield  {author} {\bibinfo {author} {\bibfnamefont {A.}~\bibnamefont
  {Lazarides}}, \bibinfo {author} {\bibfnamefont {A.}~\bibnamefont {Das}}, \
  and\ \bibinfo {author} {\bibfnamefont {R.}~\bibnamefont {Moessner}},\ }\href
  {\doibase 10.1103/PhysRevE.90.012110} {\bibfield  {journal} {\bibinfo
  {journal} {Phys. Rev. E}\ }\textbf {\bibinfo {volume} {90}},\ \bibinfo
  {pages} {012110} (\bibinfo {year} {2014}{\natexlab{a}})}\BibitemShut
  {NoStop}%
\bibitem [{\citenamefont {Ponte}\ \emph
  {et~al.}(2015{\natexlab{a}})\citenamefont {Ponte}, \citenamefont {Chandran},
  \citenamefont {Papi\'c‡},\ and\ \citenamefont {Abanin}}]{Ponte15}%
  \BibitemOpen
  \bibfield  {author} {\bibinfo {author} {\bibfnamefont {P.}~\bibnamefont
  {Ponte}}, \bibinfo {author} {\bibfnamefont {A.}~\bibnamefont {Chandran}},
  \bibinfo {author} {\bibfnamefont {Z.}~\bibnamefont {Papi\'c‡}}, \ and\
  \bibinfo {author} {\bibfnamefont {D.~A.}\ \bibnamefont {Abanin}},\ }\href
  {\doibase http://dx.doi.org/10.1016/j.aop.2014.11.008} {\bibfield  {journal}
  {\bibinfo  {journal} {Annals of Physics}\ }\textbf {\bibinfo {volume}
  {353}},\ \bibinfo {pages} {196 } (\bibinfo {year}
  {2015}{\natexlab{a}})}\BibitemShut {NoStop}%
\bibitem [{\citenamefont {Genske}\ and\ \citenamefont {Rosch}(2015)}]{Genske}%
  \BibitemOpen
  \bibfield  {author} {\bibinfo {author} {\bibfnamefont {M.}~\bibnamefont
  {Genske}}\ and\ \bibinfo {author} {\bibfnamefont {A.}~\bibnamefont {Rosch}},\
  }\href {\doibase 10.1103/PhysRevA.92.062108} {\bibfield  {journal} {\bibinfo
  {journal} {Phys. Rev. A}\ }\textbf {\bibinfo {volume} {92}},\ \bibinfo
  {pages} {062108} (\bibinfo {year} {2015})}\BibitemShut {NoStop}%
\bibitem [{\citenamefont {Abanin}\ \emph
  {et~al.}(2015{\natexlab{a}})\citenamefont {Abanin}, \citenamefont
  {De~Roeck},\ and\ \citenamefont {Huveneers}}]{DeR15}%
  \BibitemOpen
  \bibfield  {author} {\bibinfo {author} {\bibfnamefont {D.~A.}\ \bibnamefont
  {Abanin}}, \bibinfo {author} {\bibfnamefont {W.}~\bibnamefont {De~Roeck}}, \
  and\ \bibinfo {author} {\bibfnamefont {F.}~\bibnamefont {Huveneers}},\ }\href
  {\doibase 10.1103/PhysRevLett.115.256803} {\bibfield  {journal} {\bibinfo
  {journal} {Phys. Rev. Lett.}\ }\textbf {\bibinfo {volume} {115}},\ \bibinfo
  {pages} {256803} (\bibinfo {year} {2015}{\natexlab{a}})}\BibitemShut
  {NoStop}%
\bibitem [{\citenamefont {Bukov}\ \emph {et~al.}(2015)\citenamefont {Bukov},
  \citenamefont {Gopalakrishnan}, \citenamefont {Knap},\ and\ \citenamefont
  {Demler}}]{Demler15}%
  \BibitemOpen
  \bibfield  {author} {\bibinfo {author} {\bibfnamefont {M.}~\bibnamefont
  {Bukov}}, \bibinfo {author} {\bibfnamefont {S.}~\bibnamefont
  {Gopalakrishnan}}, \bibinfo {author} {\bibfnamefont {M.}~\bibnamefont
  {Knap}}, \ and\ \bibinfo {author} {\bibfnamefont {E.}~\bibnamefont
  {Demler}},\ }\href {\doibase 10.1103/PhysRevLett.115.205301} {\bibfield
  {journal} {\bibinfo  {journal} {Phys. Rev. Lett.}\ }\textbf {\bibinfo
  {volume} {115}},\ \bibinfo {pages} {205301} (\bibinfo {year}
  {2015})}\BibitemShut {NoStop}%
\bibitem [{\citenamefont {Abanin}\ \emph
  {et~al.}(2015{\natexlab{b}})\citenamefont {Abanin}, \citenamefont
  {De~Roeck},\ and\ \citenamefont {Ho}}]{Abanin15}%
  \BibitemOpen
  \bibfield  {author} {\bibinfo {author} {\bibfnamefont {D.}~\bibnamefont
  {Abanin}}, \bibinfo {author} {\bibfnamefont {W.}~\bibnamefont {De~Roeck}}, \
  and\ \bibinfo {author} {\bibfnamefont {W.~W.}\ \bibnamefont {Ho}},\
  }\href@noop {} {\bibfield  {journal} {\bibinfo  {journal} {arXiv:1510.03405}\
  } (\bibinfo {year} {2015}{\natexlab{b}})}\BibitemShut {NoStop}%
\bibitem [{\citenamefont {Amico}\ \emph {et~al.}(2008)\citenamefont {Amico},
  \citenamefont {Fazio}, \citenamefont {Osterloh},\ and\ \citenamefont
  {Vedral}}]{Saro}%
  \BibitemOpen
  \bibfield  {author} {\bibinfo {author} {\bibfnamefont {L.}~\bibnamefont
  {Amico}}, \bibinfo {author} {\bibfnamefont {R.}~\bibnamefont {Fazio}},
  \bibinfo {author} {\bibfnamefont {A.}~\bibnamefont {Osterloh}}, \ and\
  \bibinfo {author} {\bibfnamefont {V.}~\bibnamefont {Vedral}},\ }\href
  {\doibase 10.1103/RevModPhys.80.517} {\bibfield  {journal} {\bibinfo
  {journal} {Rev. Mod. Phys.}\ }\textbf {\bibinfo {volume} {80}},\ \bibinfo
  {pages} {517} (\bibinfo {year} {2008})}\BibitemShut {NoStop}%
\bibitem [{\citenamefont {Calabrese}\ \emph {et~al.}(2009)\citenamefont
  {Calabrese}, \citenamefont {Cardy},\ and\ \citenamefont {Doyon}}]{Doyon}%
  \BibitemOpen
  \bibfield  {author} {\bibinfo {author} {\bibfnamefont {P.}~\bibnamefont
  {Calabrese}}, \bibinfo {author} {\bibfnamefont {J.}~\bibnamefont {Cardy}}, \
  and\ \bibinfo {author} {\bibfnamefont {B.}~\bibnamefont {Doyon}},\
  }\href@noop {} {\bibfield  {journal} {\bibinfo  {journal} {J. Phys. A}\
  }\textbf {\bibinfo {volume} {42}} (\bibinfo {year} {2009})}\BibitemShut
  {NoStop}%
\bibitem [{\citenamefont {Its}\ \emph {et~al.}(2005)\citenamefont {Its},
  \citenamefont {Jin},\ and\ \citenamefont {Korepin}}]{0305-4470-38-13-011}%
  \BibitemOpen
  \bibfield  {author} {\bibinfo {author} {\bibfnamefont {A.~R.}\ \bibnamefont
  {Its}}, \bibinfo {author} {\bibfnamefont {B.-Q.}\ \bibnamefont {Jin}}, \ and\
  \bibinfo {author} {\bibfnamefont {V.~E.}\ \bibnamefont {Korepin}},\ }\href
  {http://stacks.iop.org/0305-4470/38/i=13/a=011} {\bibfield  {journal}
  {\bibinfo  {journal} {Journal of Physics A: Mathematical and General}\
  }\textbf {\bibinfo {volume} {38}},\ \bibinfo {pages} {2975} (\bibinfo {year}
  {2005})}\BibitemShut {NoStop}%
\bibitem [{\citenamefont {Kitaev}\ and\ \citenamefont
  {Preskill}(2006)}]{Kitaev}%
  \BibitemOpen
  \bibfield  {author} {\bibinfo {author} {\bibfnamefont {A.}~\bibnamefont
  {Kitaev}}\ and\ \bibinfo {author} {\bibfnamefont {J.}~\bibnamefont
  {Preskill}},\ }\href {\doibase 10.1103/PhysRevLett.96.110404} {\bibfield
  {journal} {\bibinfo  {journal} {Phys. Rev. Lett.}\ }\textbf {\bibinfo
  {volume} {96}},\ \bibinfo {pages} {110404} (\bibinfo {year}
  {2006})}\BibitemShut {NoStop}%
\bibitem [{\citenamefont {Levin}\ and\ \citenamefont {Wen}(2006)}]{Wen}%
  \BibitemOpen
  \bibfield  {author} {\bibinfo {author} {\bibfnamefont {M.}~\bibnamefont
  {Levin}}\ and\ \bibinfo {author} {\bibfnamefont {X.-G.}\ \bibnamefont
  {Wen}},\ }\href {\doibase 10.1103/PhysRevLett.96.110405} {\bibfield
  {journal} {\bibinfo  {journal} {Phys. Rev. Lett.}\ }\textbf {\bibinfo
  {volume} {96}},\ \bibinfo {pages} {110405} (\bibinfo {year}
  {2006})}\BibitemShut {NoStop}%
\bibitem [{\citenamefont {Jiang}\ \emph {et~al.}(2012)\citenamefont {Jiang},
  \citenamefont {Wang},\ and\ \citenamefont {Balents}}]{Jiang}%
  \BibitemOpen
  \bibfield  {author} {\bibinfo {author} {\bibfnamefont {H.-C.}\ \bibnamefont
  {Jiang}}, \bibinfo {author} {\bibfnamefont {Z.}~\bibnamefont {Wang}}, \ and\
  \bibinfo {author} {\bibfnamefont {L.}~\bibnamefont {Balents}},\ }\href@noop
  {} {\bibfield  {journal} {\bibinfo  {journal} {Nat. Phys.}\ }\textbf
  {\bibinfo {volume} {8}},\ \bibinfo {pages} {902} (\bibinfo {year}
  {2012})}\BibitemShut {NoStop}%
\bibitem [{\citenamefont {Daley}\ \emph {et~al.}(2012)\citenamefont {Daley},
  \citenamefont {Pichler}, \citenamefont {Schachenmayer},\ and\ \citenamefont
  {Zoller}}]{Daley}%
  \BibitemOpen
  \bibfield  {author} {\bibinfo {author} {\bibfnamefont {A.~J.}\ \bibnamefont
  {Daley}}, \bibinfo {author} {\bibfnamefont {H.}~\bibnamefont {Pichler}},
  \bibinfo {author} {\bibfnamefont {J.}~\bibnamefont {Schachenmayer}}, \ and\
  \bibinfo {author} {\bibfnamefont {P.}~\bibnamefont {Zoller}},\ }\href
  {\doibase 10.1103/PhysRevLett.109.020505} {\bibfield  {journal} {\bibinfo
  {journal} {Phys. Rev. Lett.}\ }\textbf {\bibinfo {volume} {109}},\ \bibinfo
  {pages} {020505} (\bibinfo {year} {2012})}\BibitemShut {NoStop}%
\bibitem [{\citenamefont {Lubasch}\ \emph {et~al.}(2011)\citenamefont
  {Lubasch}, \citenamefont {Mintert},\ and\ \citenamefont
  {Wimberger}}]{PhysRevA.84.063615}%
  \BibitemOpen
  \bibfield  {author} {\bibinfo {author} {\bibfnamefont {M.}~\bibnamefont
  {Lubasch}}, \bibinfo {author} {\bibfnamefont {F.}~\bibnamefont {Mintert}}, \
  and\ \bibinfo {author} {\bibfnamefont {S.}~\bibnamefont {Wimberger}},\ }\href
  {\doibase 10.1103/PhysRevA.84.063615} {\bibfield  {journal} {\bibinfo
  {journal} {Phys. Rev. A}\ }\textbf {\bibinfo {volume} {84}},\ \bibinfo
  {pages} {063615} (\bibinfo {year} {2011})}\BibitemShut {NoStop}%
\bibitem [{\citenamefont {Gori}\ \emph {et~al.}(2015)\citenamefont {Gori},
  \citenamefont {Paganelli}, \citenamefont {Sharma}, \citenamefont {Sodano},\
  and\ \citenamefont {Trombettoni}}]{PhysRevB.91.245138}%
  \BibitemOpen
  \bibfield  {author} {\bibinfo {author} {\bibfnamefont {G.}~\bibnamefont
  {Gori}}, \bibinfo {author} {\bibfnamefont {S.}~\bibnamefont {Paganelli}},
  \bibinfo {author} {\bibfnamefont {A.}~\bibnamefont {Sharma}}, \bibinfo
  {author} {\bibfnamefont {P.}~\bibnamefont {Sodano}}, \ and\ \bibinfo {author}
  {\bibfnamefont {A.}~\bibnamefont {Trombettoni}},\ }\href {\doibase
  10.1103/PhysRevB.91.245138} {\bibfield  {journal} {\bibinfo  {journal} {Phys.
  Rev. B}\ }\textbf {\bibinfo {volume} {91}},\ \bibinfo {pages} {245138}
  (\bibinfo {year} {2015})}\BibitemShut {NoStop}%
\bibitem [{\citenamefont {Sen}\ and\ \citenamefont {Sengupta}(2015)}]{Sen}%
  \BibitemOpen
  \bibfield  {author} {\bibinfo {author} {\bibfnamefont {A.}~\bibnamefont
  {Sen}}\ and\ \bibinfo {author} {\bibfnamefont {K.}~\bibnamefont {Sengupta}},\
  }\href@noop {} {\bibfield  {journal} {\bibinfo  {journal} {arXiv:1511.03668}\
  } (\bibinfo {year} {2015})}\BibitemShut {NoStop}%
\bibitem [{\citenamefont {Ponte}\ \emph
  {et~al.}(2015{\natexlab{b}})\citenamefont {Ponte}, \citenamefont
  {Papi\ifmmode~\acute{c}\else \'{c}\fi{}}, \citenamefont {Huveneers},\ and\
  \citenamefont {Abanin}}]{CC}%
  \BibitemOpen
  \bibfield  {author} {\bibinfo {author} {\bibfnamefont {P.}~\bibnamefont
  {Ponte}}, \bibinfo {author} {\bibfnamefont {Z.}~\bibnamefont
  {Papi\ifmmode~\acute{c}\else \'{c}\fi{}}}, \bibinfo {author} {\bibfnamefont
  {F.}~\bibnamefont {Huveneers}}, \ and\ \bibinfo {author} {\bibfnamefont
  {D.~A.}\ \bibnamefont {Abanin}},\ }\href {\doibase
  10.1103/PhysRevLett.114.140401} {\bibfield  {journal} {\bibinfo  {journal}
  {Phys. Rev. Lett.}\ }\textbf {\bibinfo {volume} {114}},\ \bibinfo {pages}
  {140401} (\bibinfo {year} {2015}{\natexlab{b}})}\BibitemShut {NoStop}%
\bibitem [{\citenamefont {Eisler}\ and\ \citenamefont
  {Peschel}(2008)}]{ANDP:ANDP200810299}%
  \BibitemOpen
  \bibfield  {author} {\bibinfo {author} {\bibfnamefont {V.}~\bibnamefont
  {Eisler}}\ and\ \bibinfo {author} {\bibfnamefont {I.}~\bibnamefont
  {Peschel}},\ }\href {\doibase 10.1002/andp.200810299} {\bibfield  {journal}
  {\bibinfo  {journal} {Annalen der Physik}\ }\textbf {\bibinfo {volume}
  {17}},\ \bibinfo {pages} {410} (\bibinfo {year} {2008})}\BibitemShut
  {NoStop}%
\bibitem [{\citenamefont {Peschel}\ and\ \citenamefont
  {Eisler}(2009)}]{1751-8121-42-50-504003}%
  \BibitemOpen
  \bibfield  {author} {\bibinfo {author} {\bibfnamefont {I.}~\bibnamefont
  {Peschel}}\ and\ \bibinfo {author} {\bibfnamefont {V.}~\bibnamefont
  {Eisler}},\ }\href {http://stacks.iop.org/1751-8121/42/i=50/a=504003}
  {\bibfield  {journal} {\bibinfo  {journal} {Journal of Physics A:
  Mathematical and Theoretical}\ }\textbf {\bibinfo {volume} {42}},\ \bibinfo
  {pages} {504003} (\bibinfo {year} {2009})}\BibitemShut {NoStop}%
\bibitem [{\citenamefont {Mishra}\ \emph {et~al.}(2015)\citenamefont {Mishra},
  \citenamefont {Lakshminarayan},\ and\ \citenamefont
  {Subrahmanyam}}]{PhysRevA.91.022318}%
  \BibitemOpen
  \bibfield  {author} {\bibinfo {author} {\bibfnamefont {S.~K.}\ \bibnamefont
  {Mishra}}, \bibinfo {author} {\bibfnamefont {A.}~\bibnamefont
  {Lakshminarayan}}, \ and\ \bibinfo {author} {\bibfnamefont {V.}~\bibnamefont
  {Subrahmanyam}},\ }\href {\doibase 10.1103/PhysRevA.91.022318} {\bibfield
  {journal} {\bibinfo  {journal} {Phys. Rev. A}\ }\textbf {\bibinfo {volume}
  {91}},\ \bibinfo {pages} {022318} (\bibinfo {year} {2015})}\BibitemShut
  {NoStop}%
\bibitem [{\citenamefont {Barmettler}\ \emph {et~al.}(2008)\citenamefont
  {Barmettler}, \citenamefont {Rey}, \citenamefont {Demler}, \citenamefont
  {Lukin}, \citenamefont {Bloch},\ and\ \citenamefont
  {Gritsev}}]{PhysRevA.78.012330}%
  \BibitemOpen
  \bibfield  {author} {\bibinfo {author} {\bibfnamefont {P.}~\bibnamefont
  {Barmettler}}, \bibinfo {author} {\bibfnamefont {A.~M.}\ \bibnamefont {Rey}},
  \bibinfo {author} {\bibfnamefont {E.}~\bibnamefont {Demler}}, \bibinfo
  {author} {\bibfnamefont {M.~D.}\ \bibnamefont {Lukin}}, \bibinfo {author}
  {\bibfnamefont {I.}~\bibnamefont {Bloch}}, \ and\ \bibinfo {author}
  {\bibfnamefont {V.}~\bibnamefont {Gritsev}},\ }\href {\doibase
  10.1103/PhysRevA.78.012330} {\bibfield  {journal} {\bibinfo  {journal} {Phys.
  Rev. A}\ }\textbf {\bibinfo {volume} {78}},\ \bibinfo {pages} {012330}
  (\bibinfo {year} {2008})}\BibitemShut {NoStop}%
\bibitem [{\citenamefont {Fagotti}\ and\ \citenamefont
  {Calabrese}(2008)}]{Fagotti08}%
  \BibitemOpen
  \bibfield  {author} {\bibinfo {author} {\bibfnamefont {M.}~\bibnamefont
  {Fagotti}}\ and\ \bibinfo {author} {\bibfnamefont {P.}~\bibnamefont
  {Calabrese}},\ }\href {\doibase 10.1103/PhysRevA.78.010306} {\bibfield
  {journal} {\bibinfo  {journal} {Phys. Rev. A}\ }\textbf {\bibinfo {volume}
  {78}},\ \bibinfo {pages} {010306} (\bibinfo {year} {2008})}\BibitemShut
  {NoStop}%
\bibitem [{\citenamefont {L\"auchli}\ and\ \citenamefont
  {Kollath}(2009)}]{Kollath}%
  \BibitemOpen
  \bibfield  {author} {\bibinfo {author} {\bibfnamefont {A.}~\bibnamefont
  {L\"auchli}}\ and\ \bibinfo {author} {\bibfnamefont {C.}~\bibnamefont
  {Kollath}},\ }\href@noop {} {\bibfield  {journal} {\bibinfo  {journal} {J.
  Stat. Mech.}\ ,\ \bibinfo {pages} {P05018}} (\bibinfo {year}
  {2009})}\BibitemShut {NoStop}%
\bibitem [{\citenamefont {Chiara}\ \emph {et~al.}(2006)\citenamefont {Chiara},
  \citenamefont {Montangero}, \citenamefont {Calabrese},\ and\ \citenamefont
  {Fazio}}]{DeChiara}%
  \BibitemOpen
  \bibfield  {author} {\bibinfo {author} {\bibfnamefont {G.~D.}\ \bibnamefont
  {Chiara}}, \bibinfo {author} {\bibfnamefont {S.}~\bibnamefont {Montangero}},
  \bibinfo {author} {\bibfnamefont {P.}~\bibnamefont {Calabrese}}, \ and\
  \bibinfo {author} {\bibfnamefont {R.}~\bibnamefont {Fazio}},\ }\href
  {http://stacks.iop.org/1742-5468/2006/i=03/a=P03001} {\bibfield  {journal}
  {\bibinfo  {journal} {Journal of Statistical Mechanics: Theory and
  Experiment}\ }\textbf {\bibinfo {volume} {2006}},\ \bibinfo {pages} {P03001}
  (\bibinfo {year} {2006})}\BibitemShut {NoStop}%
\bibitem [{\citenamefont {Igl\'oi}\ \emph {et~al.}(2012)\citenamefont
  {Igl\'oi}, \citenamefont {Szatm\'ari},\ and\ \citenamefont {Lin}}]{Igloi}%
  \BibitemOpen
  \bibfield  {author} {\bibinfo {author} {\bibfnamefont {F.}~\bibnamefont
  {Igl\'oi}}, \bibinfo {author} {\bibfnamefont {Z.}~\bibnamefont {Szatm\'ari}},
  \ and\ \bibinfo {author} {\bibfnamefont {Y.-C.}\ \bibnamefont {Lin}},\ }\href
  {\doibase 10.1103/PhysRevB.85.094417} {\bibfield  {journal} {\bibinfo
  {journal} {Phys. Rev. B}\ }\textbf {\bibinfo {volume} {85}},\ \bibinfo
  {pages} {094417} (\bibinfo {year} {2012})}\BibitemShut {NoStop}%
\bibitem [{\citenamefont {Fagotti}\ and\ \citenamefont
  {Collura}(2015)}]{Collura}%
  \BibitemOpen
  \bibfield  {author} {\bibinfo {author} {\bibfnamefont {M.}~\bibnamefont
  {Fagotti}}\ and\ \bibinfo {author} {\bibfnamefont {M.}~\bibnamefont
  {Collura}},\ }\href@noop {} {\bibfield  {journal} {\bibinfo  {journal}
  {arXiv:1507.02678}\ } (\bibinfo {year} {2015})}\BibitemShut {NoStop}%
\bibitem [{\citenamefont {{Pitsios}}\ \emph {et~al.}(2016)\citenamefont
  {{Pitsios}}, \citenamefont {{Banchi}}, \citenamefont {{Rab}}, \citenamefont
  {{Bentivegna}}, \citenamefont {{Caprara}}, \citenamefont {{Crespi}},
  \citenamefont {{Spagnolo}}, \citenamefont {{Bose}}, \citenamefont
  {{Mataloni}}, \citenamefont {{Osellame}},\ and\ \citenamefont
  {{Sciarrino}}}]{2016arXiv160302669P}%
  \BibitemOpen
  \bibfield  {author} {\bibinfo {author} {\bibfnamefont {I.}~\bibnamefont
  {{Pitsios}}}, \bibinfo {author} {\bibfnamefont {L.}~\bibnamefont {{Banchi}}},
  \bibinfo {author} {\bibfnamefont {A.~S.}\ \bibnamefont {{Rab}}}, \bibinfo
  {author} {\bibfnamefont {M.}~\bibnamefont {{Bentivegna}}}, \bibinfo {author}
  {\bibfnamefont {D.}~\bibnamefont {{Caprara}}}, \bibinfo {author}
  {\bibfnamefont {A.}~\bibnamefont {{Crespi}}}, \bibinfo {author}
  {\bibfnamefont {N.}~\bibnamefont {{Spagnolo}}}, \bibinfo {author}
  {\bibfnamefont {S.}~\bibnamefont {{Bose}}}, \bibinfo {author} {\bibfnamefont
  {P.}~\bibnamefont {{Mataloni}}}, \bibinfo {author} {\bibfnamefont
  {R.}~\bibnamefont {{Osellame}}}, \ and\ \bibinfo {author} {\bibfnamefont
  {F.}~\bibnamefont {{Sciarrino}}},\ }\href@noop {} {\bibfield  {journal}
  {\bibinfo  {journal} {ArXiv e-prints}\ } (\bibinfo {year} {2016})},\ \Eprint
  {http://arxiv.org/abs/1603.02669} {arXiv:1603.02669 [quant-ph]} \BibitemShut
  {NoStop}%
\bibitem [{\citenamefont {Canovi}\ \emph {et~al.}(2014)\citenamefont {Canovi},
  \citenamefont {Ercolessi}, \citenamefont {Naldesi}, \citenamefont {Taddia},\
  and\ \citenamefont {Vodola}}]{PhysRevB.89.104303}%
  \BibitemOpen
  \bibfield  {author} {\bibinfo {author} {\bibfnamefont {E.}~\bibnamefont
  {Canovi}}, \bibinfo {author} {\bibfnamefont {E.}~\bibnamefont {Ercolessi}},
  \bibinfo {author} {\bibfnamefont {P.}~\bibnamefont {Naldesi}}, \bibinfo
  {author} {\bibfnamefont {L.}~\bibnamefont {Taddia}}, \ and\ \bibinfo {author}
  {\bibfnamefont {D.}~\bibnamefont {Vodola}},\ }\href {\doibase
  10.1103/PhysRevB.89.104303} {\bibfield  {journal} {\bibinfo  {journal} {Phys.
  Rev. B}\ }\textbf {\bibinfo {volume} {89}},\ \bibinfo {pages} {104303}
  (\bibinfo {year} {2014})}\BibitemShut {NoStop}%
\bibitem [{\citenamefont {Russomanno}\ \emph {et~al.}(2012)\citenamefont
  {Russomanno}, \citenamefont {Silva},\ and\ \citenamefont {Santoro}}]{Angelo}%
  \BibitemOpen
  \bibfield  {author} {\bibinfo {author} {\bibfnamefont {A.}~\bibnamefont
  {Russomanno}}, \bibinfo {author} {\bibfnamefont {A.}~\bibnamefont {Silva}}, \
  and\ \bibinfo {author} {\bibfnamefont {G.~E.}\ \bibnamefont {Santoro}},\
  }\href {\doibase 10.1103/PhysRevLett.109.257201} {\bibfield  {journal}
  {\bibinfo  {journal} {Phys. Rev. Lett.}\ }\textbf {\bibinfo {volume} {109}},\
  \bibinfo {pages} {257201} (\bibinfo {year} {2012})}\BibitemShut {NoStop}%
\bibitem [{\citenamefont {Russomanno}\ \emph {et~al.}(2013)\citenamefont
  {Russomanno}, \citenamefont {Silva},\ and\ \citenamefont
  {Santoro}}]{Angelo2}%
  \BibitemOpen
  \bibfield  {author} {\bibinfo {author} {\bibfnamefont {A.}~\bibnamefont
  {Russomanno}}, \bibinfo {author} {\bibfnamefont {A.}~\bibnamefont {Silva}}, \
  and\ \bibinfo {author} {\bibfnamefont {G.~E.}\ \bibnamefont {Santoro}},\
  }\href@noop {} {\bibfield  {journal} {\bibinfo  {journal} {J. Stat. Mech.}\
  ,\ \bibinfo {pages} {P09012}} (\bibinfo {year} {2013})}\BibitemShut {NoStop}%
\bibitem [{\citenamefont {Lazarides}\ \emph
  {et~al.}(2014{\natexlab{b}})\citenamefont {Lazarides}, \citenamefont {Das},\
  and\ \citenamefont {Moessner}}]{Laz}%
  \BibitemOpen
  \bibfield  {author} {\bibinfo {author} {\bibfnamefont {A.}~\bibnamefont
  {Lazarides}}, \bibinfo {author} {\bibfnamefont {A.}~\bibnamefont {Das}}, \
  and\ \bibinfo {author} {\bibfnamefont {R.}~\bibnamefont {Moessner}},\ }\href
  {\doibase 10.1103/PhysRevLett.112.150401} {\bibfield  {journal} {\bibinfo
  {journal} {Phys. Rev. Lett.}\ }\textbf {\bibinfo {volume} {112}},\ \bibinfo
  {pages} {150401} (\bibinfo {year} {2014}{\natexlab{b}})}\BibitemShut
  {NoStop}%
\bibitem [{\citenamefont {Roy}\ and\ \citenamefont {Das}(2015)}]{Roy}%
  \BibitemOpen
  \bibfield  {author} {\bibinfo {author} {\bibfnamefont {A.}~\bibnamefont
  {Roy}}\ and\ \bibinfo {author} {\bibfnamefont {A.}~\bibnamefont {Das}},\
  }\href {\doibase 10.1103/PhysRevB.91.121106} {\bibfield  {journal} {\bibinfo
  {journal} {Phys. Rev. B}\ }\textbf {\bibinfo {volume} {91}},\ \bibinfo
  {pages} {121106} (\bibinfo {year} {2015})}\BibitemShut {NoStop}%
\bibitem [{\citenamefont {Lazarides}\ \emph {et~al.}(2015)\citenamefont
  {Lazarides}, \citenamefont {Das},\ and\ \citenamefont {Moessner}}]{LazMBL}%
  \BibitemOpen
  \bibfield  {author} {\bibinfo {author} {\bibfnamefont {A.}~\bibnamefont
  {Lazarides}}, \bibinfo {author} {\bibfnamefont {A.}~\bibnamefont {Das}}, \
  and\ \bibinfo {author} {\bibfnamefont {R.}~\bibnamefont {Moessner}},\ }\href
  {\doibase 10.1103/PhysRevLett.115.030402} {\bibfield  {journal} {\bibinfo
  {journal} {Phys. Rev. Lett.}\ }\textbf {\bibinfo {volume} {115}},\ \bibinfo
  {pages} {030402} (\bibinfo {year} {2015})}\BibitemShut {NoStop}%
\bibitem [{\citenamefont {Sachdev}(1999)}]{Sachdev}%
  \BibitemOpen
  \bibfield  {author} {\bibinfo {author} {\bibfnamefont {S.}~\bibnamefont
  {Sachdev}},\ }\href@noop {} {\emph {\bibinfo {title} {Quantum Phase
  Transitions}}}\ (\bibinfo  {publisher} {Cambridge: Cambridge University
  Press},\ \bibinfo {year} {1999})\BibitemShut {NoStop}%
\bibitem [{\citenamefont {De~Pasquale}\ and\ \citenamefont
  {Facchi}(2009)}]{PhysRevA.80.032102}%
  \BibitemOpen
  \bibfield  {author} {\bibinfo {author} {\bibfnamefont {A.}~\bibnamefont
  {De~Pasquale}}\ and\ \bibinfo {author} {\bibfnamefont {P.}~\bibnamefont
  {Facchi}},\ }\href {\doibase 10.1103/PhysRevA.80.032102} {\bibfield
  {journal} {\bibinfo  {journal} {Phys. Rev. A}\ }\textbf {\bibinfo {volume}
  {80}},\ \bibinfo {pages} {032102} (\bibinfo {year} {2009})}\BibitemShut
  {NoStop}%
\bibitem [{\citenamefont {Damski}\ and\ \citenamefont
  {Rams}(2014)}]{1751-8121-47-2-025303}%
  \BibitemOpen
  \bibfield  {author} {\bibinfo {author} {\bibfnamefont {B.}~\bibnamefont
  {Damski}}\ and\ \bibinfo {author} {\bibfnamefont {M.~M.}\ \bibnamefont
  {Rams}},\ }\href {http://stacks.iop.org/1751-8121/47/i=2/a=025303} {\bibfield
   {journal} {\bibinfo  {journal} {Journal of Physics A: Mathematical and
  Theoretical}\ }\textbf {\bibinfo {volume} {47}},\ \bibinfo {pages} {025303}
  (\bibinfo {year} {2014})}\BibitemShut {NoStop}%
\bibitem [{\citenamefont {Caux}(2016)}]{Caux}%
  \BibitemOpen
  \bibfield  {author} {\bibinfo {author} {\bibfnamefont {J.~S.}\ \bibnamefont
  {Caux}},\ }\href@noop {} {\bibfield  {journal} {\bibinfo  {journal}
  {arXiv:1603.04689v1}\ } (\bibinfo {year} {2016})}\BibitemShut {NoStop}%
\bibitem [{\citenamefont {Essler}\ and\ \citenamefont
  {Fagotti}(2016)}]{Essler}%
  \BibitemOpen
  \bibfield  {author} {\bibinfo {author} {\bibfnamefont {F.~H.~L.}\
  \bibnamefont {Essler}}\ and\ \bibinfo {author} {\bibfnamefont
  {M.}~\bibnamefont {Fagotti}},\ }\href@noop {} {\bibfield  {journal} {\bibinfo
   {journal} {arXiv:1603.06452v1}\ } (\bibinfo {year} {2016})}\BibitemShut
  {NoStop}%
\bibitem [{\citenamefont {Peschel}(2004)}]{1742-5468-2004-12-P12005}%
  \BibitemOpen
  \bibfield  {author} {\bibinfo {author} {\bibfnamefont {I.}~\bibnamefont
  {Peschel}},\ }\href {http://stacks.iop.org/1742-5468/2004/i=12/a=P12005}
  {\bibfield  {journal} {\bibinfo  {journal} {Journal of Statistical Mechanics:
  Theory and Experiment}\ }\textbf {\bibinfo {volume} {2004}},\ \bibinfo
  {pages} {P12005} (\bibinfo {year} {2004})}\BibitemShut {NoStop}%
\bibitem [{\citenamefont {Franchini}\ \emph {et~al.}(2008)\citenamefont
  {Franchini}, \citenamefont {Its},\ and\ \citenamefont
  {Korepin}}]{FranchiniJPA08}%
  \BibitemOpen
  \bibfield  {author} {\bibinfo {author} {\bibfnamefont {F.}~\bibnamefont
  {Franchini}}, \bibinfo {author} {\bibfnamefont {A.~R.}\ \bibnamefont {Its}},
  \ and\ \bibinfo {author} {\bibfnamefont {V.~E.}\ \bibnamefont {Korepin}},\
  }\href {http://stacks.iop.org/1751-8121/41/i=2/a=025302} {\bibfield
  {journal} {\bibinfo  {journal} {Journal of Physics A: Mathematical and
  Theoretical}\ }\textbf {\bibinfo {volume} {41}},\ \bibinfo {pages} {025302}
  (\bibinfo {year} {2008})}\BibitemShut {NoStop}%
\bibitem [{\citenamefont {Franchini}\ \emph {et~al.}(2007)\citenamefont
  {Franchini}, \citenamefont {Its}, \citenamefont {Jin},\ and\ \citenamefont
  {Korepin}}]{FranchiniJPA07}%
  \BibitemOpen
  \bibfield  {author} {\bibinfo {author} {\bibfnamefont {F.}~\bibnamefont
  {Franchini}}, \bibinfo {author} {\bibfnamefont {A.~R.}\ \bibnamefont {Its}},
  \bibinfo {author} {\bibfnamefont {B.-Q.}\ \bibnamefont {Jin}}, \ and\
  \bibinfo {author} {\bibfnamefont {V.~E.}\ \bibnamefont {Korepin}},\ }\href
  {http://stacks.iop.org/1751-8121/40/i=29/a=019} {\bibfield  {journal}
  {\bibinfo  {journal} {Journal of Physics A: Mathematical and Theoretical}\
  }\textbf {\bibinfo {volume} {40}},\ \bibinfo {pages} {8467} (\bibinfo {year}
  {2007})}\BibitemShut {NoStop}%
\bibitem [{\citenamefont {Blass}\ \emph {et~al.}(2012)\citenamefont {Blass},
  \citenamefont {Rieger},\ and\ \citenamefont {Igloi}}]{igloi2}%
  \BibitemOpen
  \bibfield  {author} {\bibinfo {author} {\bibfnamefont {B.}~\bibnamefont
  {Blass}}, \bibinfo {author} {\bibfnamefont {H.}~\bibnamefont {Rieger}}, \
  and\ \bibinfo {author} {\bibfnamefont {F.}~\bibnamefont {Igloi}},\
  }\href@noop {} {\bibfield  {journal} {\bibinfo  {journal} {Europhys. Lett.}\
  }\textbf {\bibinfo {volume} {99}},\ \bibinfo {pages} {041050} (\bibinfo
  {year} {2012})}\BibitemShut {NoStop}%
\bibitem [{Note1()}]{Note1}%
  \BibitemOpen
  \bibinfo {note} {With the notable exception of the critical point where it
  scales logarithmically $S_l\sim \protect \qopname \relax o{log}l$, see for
  instance \cite {Doyon, Saro}.}\BibitemShut {Stop}%
\bibitem [{\citenamefont {Vidal}\ \emph {et~al.}(2003)\citenamefont {Vidal},
  \citenamefont {Latorre}, \citenamefont {Rico},\ and\ \citenamefont
  {Kitaev}}]{Vidal2003}%
  \BibitemOpen
  \bibfield  {author} {\bibinfo {author} {\bibfnamefont {G.}~\bibnamefont
  {Vidal}}, \bibinfo {author} {\bibfnamefont {J.~I.}\ \bibnamefont {Latorre}},
  \bibinfo {author} {\bibfnamefont {E.}~\bibnamefont {Rico}}, \ and\ \bibinfo
  {author} {\bibfnamefont {A.}~\bibnamefont {Kitaev}},\ }\href {\doibase
  10.1103/PhysRevLett.90.227902} {\bibfield  {journal} {\bibinfo  {journal}
  {Phys. Rev. Lett.}\ }\textbf {\bibinfo {volume} {90}},\ \bibinfo {pages}
  {227902} (\bibinfo {year} {2003})}\BibitemShut {NoStop}%
\bibitem [{\citenamefont {Cincio}\ \emph {et~al.}(2007)\citenamefont {Cincio},
  \citenamefont {Dziarmaga}, \citenamefont {Rams},\ and\ \citenamefont
  {Zurek}}]{Cincio2007}%
  \BibitemOpen
  \bibfield  {author} {\bibinfo {author} {\bibfnamefont {L.}~\bibnamefont
  {Cincio}}, \bibinfo {author} {\bibfnamefont {J.}~\bibnamefont {Dziarmaga}},
  \bibinfo {author} {\bibfnamefont {M.~M.}\ \bibnamefont {Rams}}, \ and\
  \bibinfo {author} {\bibfnamefont {W.~H.}\ \bibnamefont {Zurek}},\ }\href
  {\doibase 10.1103/PhysRevA.75.052321} {\bibfield  {journal} {\bibinfo
  {journal} {Phys. Rev. A}\ }\textbf {\bibinfo {volume} {75}},\ \bibinfo
  {pages} {052321} (\bibinfo {year} {2007})}\BibitemShut {NoStop}%
\bibitem [{\citenamefont {Latorre}\ and\ \citenamefont
  {Riera}(2009)}]{latorre}%
  \BibitemOpen
  \bibfield  {author} {\bibinfo {author} {\bibfnamefont {J.~I.}\ \bibnamefont
  {Latorre}}\ and\ \bibinfo {author} {\bibfnamefont {A.}~\bibnamefont
  {Riera}},\ }\href {http://stacks.iop.org/1751-8121/42/i=50/a=504002}
  {\bibfield  {journal} {\bibinfo  {journal} {Journal of Physics A:
  Mathematical and Theoretical}\ }\textbf {\bibinfo {volume} {42}},\ \bibinfo
  {pages} {504002} (\bibinfo {year} {2009})}\BibitemShut {NoStop}%
\bibitem [{\citenamefont {Calabrese}\ \emph {et~al.}(2012)\citenamefont
  {Calabrese}, \citenamefont {Essler},\ and\ \citenamefont
  {Fagotti}}]{1742-5468-2012-07-P07016}%
  \BibitemOpen
  \bibfield  {author} {\bibinfo {author} {\bibfnamefont {P.}~\bibnamefont
  {Calabrese}}, \bibinfo {author} {\bibfnamefont {F.~H.~L.}\ \bibnamefont
  {Essler}}, \ and\ \bibinfo {author} {\bibfnamefont {M.}~\bibnamefont
  {Fagotti}},\ }\href {http://stacks.iop.org/1742-5468/2012/i=07/a=P07016}
  {\bibfield  {journal} {\bibinfo  {journal} {Journal of Statistical Mechanics:
  Theory and Experiment}\ }\textbf {\bibinfo {volume} {2012}},\ \bibinfo
  {pages} {P07016} (\bibinfo {year} {2012})}\BibitemShut {NoStop}%
\bibitem [{\citenamefont {Johansson}\ \emph {et~al.}(2013)\citenamefont
  {Johansson}, \citenamefont {Nation},\ and\ \citenamefont {Nori}}]{Nori}%
  \BibitemOpen
  \bibfield  {author} {\bibinfo {author} {\bibfnamefont {J.}~\bibnamefont
  {Johansson}}, \bibinfo {author} {\bibfnamefont {P.}~\bibnamefont {Nation}}, \
  and\ \bibinfo {author} {\bibfnamefont {F.}~\bibnamefont {Nori}},\ }\href@noop
  {} {\bibfield  {journal} {\bibinfo  {journal} {Computer Physics
  Communications}\ }\textbf {\bibinfo {volume} {184}},\ \bibinfo {pages} {1234
  } (\bibinfo {year} {2013})}\BibitemShut {NoStop}%
\bibitem [{Note2()}]{Note2}%
  \BibitemOpen
  \bibinfo {note} {As a matter of fact, the only exception is represented by
  drivings starting at $h(t=0)=0$; in this case entanglement entropy oscillates
  with frequency $2\omega $.}\BibitemShut {Stop}%
\bibitem [{Note3()}]{Note3}%
  \BibitemOpen
  \bibinfo {note} {The driving $h(t)=2+\protect \frac {1}{2}\protect \qopname
  \relax o{sin}\protect \frac {\pi }{10}t$ has not been included in Fig. \ref
  {esse} as $S_l(nT)$ exhibits a relaxation time beyond our numerical
  investigation and, thus, no plateaux appears in the dynamics of
  $S_l(t)$}\BibitemShut {NoStop}%
\bibitem [{\citenamefont {Igl\'{o}i}\ and\ \citenamefont
  {Lin}(2008)}]{IgloiLinJSM08}%
  \BibitemOpen
  \bibfield  {author} {\bibinfo {author} {\bibfnamefont {F.}~\bibnamefont
  {Igl\'{o}i}}\ and\ \bibinfo {author} {\bibfnamefont {Y.-C.}\ \bibnamefont
  {Lin}},\ }\href {http://stacks.iop.org/1742-5468/2008/i=06/a=P06004}
  {\bibfield  {journal} {\bibinfo  {journal} {Journal of Statistical Mechanics:
  Theory and Experiment}\ }\textbf {\bibinfo {volume} {2008}},\ \bibinfo
  {pages} {P06004} (\bibinfo {year} {2008})}\BibitemShut {NoStop}%
\bibitem [{\citenamefont {Sondhi}\ \emph {et~al.}(1997)\citenamefont {Sondhi},
  \citenamefont {Girvin}, \citenamefont {Carini},\ and\ \citenamefont
  {Shahar}}]{Sondhi}%
  \BibitemOpen
  \bibfield  {author} {\bibinfo {author} {\bibfnamefont {S.~L.}\ \bibnamefont
  {Sondhi}}, \bibinfo {author} {\bibfnamefont {S.~M.}\ \bibnamefont {Girvin}},
  \bibinfo {author} {\bibfnamefont {J.~P.}\ \bibnamefont {Carini}}, \ and\
  \bibinfo {author} {\bibfnamefont {D.}~\bibnamefont {Shahar}},\ }\href
  {\doibase 10.1103/RevModPhys.69.315} {\bibfield  {journal} {\bibinfo
  {journal} {Rev. Mod. Phys.}\ }\textbf {\bibinfo {volume} {69}},\ \bibinfo
  {pages} {315} (\bibinfo {year} {1997})}\BibitemShut {NoStop}%
\bibitem [{\citenamefont {Russomanno}\ \emph {et~al.}(2016)\citenamefont
  {Russomanno}, \citenamefont {Santoro},\ and\ \citenamefont
  {Fazio}}]{AngeloNEW}%
  \BibitemOpen
  \bibfield  {author} {\bibinfo {author} {\bibfnamefont {A.}~\bibnamefont
  {Russomanno}}, \bibinfo {author} {\bibfnamefont {G.~E.}\ \bibnamefont
  {Santoro}}, \ and\ \bibinfo {author} {\bibfnamefont {R.}~\bibnamefont
  {Fazio}},\ }\href@noop {} {\bibfield  {journal} {\bibinfo  {journal}
  {arXiv:1603.03663}\ } (\bibinfo {year} {2016})}\BibitemShut {NoStop}%
\end{thebibliography}%

\end{document}